\documentclass[nonacm,sigconf]{acmart}

\renewcommand\footnotetextcopyrightpermission[1]{}
\AtBeginDocument{%
  }

\usepackage{todonotes}
\usepackage{multirow}

\usepackage{natbib}
\usepackage{caption}

\usepackage[vlined,linesnumbered,ruled,resetcount]{algorithm2e}
\usepackage{algorithmicx, algpseudocode}
\usepackage{amsmath}
\usepackage{enumitem}
\usepackage{newfloat}
\usepackage{listings}
\usepackage[many]{tcolorbox}

\definecolor{lightgray}{rgb}{.9,.9,.9}
\definecolor{snow}{rgb}{0.96, 0.95, 0.95}
\definecolor{darkgray}{rgb}{.4,.4,.4}
\definecolor{applegreen}{rgb}{0.55, 0.71, 0.0}
\definecolor{forestgreen}{rgb}{0.13, 0.55, 0.13}
\definecolor{purple}{rgb}{0.65, 0.12, 0.82}

\usepackage{makecell}
\usepackage{graphicx}

\usepackage{float}
\floatstyle{plaintop}
\newfloat{lstfloat}{tbp}{lop}
\floatname{lstfloat}{Listing}

\usepackage{pgfplots}
\usepackage{subcaption}
\pgfplotsset{compat=1.17}

\newcommand{\ourfuzzer}{CovRL-Fuzz}
\newcommand{\javascript}{JS}
\newcommand*\circled[1]{\tikz[baseline=(char.base)]{
            \node[shape=circle,fill,inner sep=0.2pt] (char) {\textcolor{white}{#1}};}}
\newcommand{\CovRL}{CovRL}
\newcommand*\circledL[1]{\tikz[baseline=-0.65ex]{
            \node[shape=circle,fill,inner sep=2.4pt] (char) {\textcolor{white}{#1}};}}

\lstdefinelanguage{JavaScript}{
  keywords={typeof, new, true, false, catch, function, class, extend, static, return, null, catch, switch, var, if, in, while, do, else, case, break},
  keywordstyle=\color{forestgreen}\bfseries,
  keywords=[2]{int ,float },
  keywordstyle=[2]\color{blue}\bfseries,
  ndkeywords={class, export, boolean, throw, implements, import, this},
  ndkeywordstyle=\color{darkgray}\bfseries,
  identifierstyle=\color{black},
  sensitive=false,
  comment=[l]{//},
  numbers=left, 
  frame=topbot, 
  numbersep=0pt, 
  captionpos=b,
  morecomment=[s]{/*}{*/},
  basicstyle=\small\ttfamily,
  commentstyle=\color{purple}\ttfamily,
  stringstyle=\color{blue}\ttfamily,
  numberstyle=\color{darkgray}\ttfamily,
  morestring=[b]',
  morestring=[b]"
}

\usepackage{pifont}
\newcommand{\cmark}{\ding{51}}%

\newcommand{\bugTotal}{48}
\newcommand{\bugCve}{11}
\newcommand{\bugUndiscover}{39}

\usepackage{colortbl}
\definecolor{lightgray2}{gray}{0.95}

\usepackage{enumitem}

\begin{document}

\title{CovRL: Fuzzing JavaScript Engines with Coverage-Guided Reinforcement Learning for LLM-based Mutation}

\author{Jueon Eom}
\affiliation{%
  \institution{Yonsei University}
  \city{}
  \country{}
}
\email{jueoneom@yonsei.ac.kr}

\author{Seyeon Jeong}
\affiliation{%
  \institution{Suresofttech Inc.}
  \city{}
  \country{}
}
\email{best6653@gmail.com}

\author{Taekyoung Kwon}
\affiliation{%
  \institution{Yonsei University}
  \city{}
  \country{}
}
\email{taekyoung@yonsei.ac.kr}

\begin{abstract}
Fuzzing is an effective bug-finding technique but it struggles with complex systems like JavaScript engines that demand precise grammatical input. Recently, researchers have adopted language models for context-aware mutation in fuzzing to address this problem. However, existing techniques are limited in utilizing coverage guidance for fuzzing, which is rather performed in a black-box manner.

This paper presents a novel technique called CovRL (Coverage-guided Reinforcement Learning) that combines Large Language Models (LLMs) with reinforcement learning from coverage feedback. Our fuzzer, CovRL-Fuzz, integrates coverage feedback directly into the LLM by leveraging the Term Frequency-Inverse Document Frequency (TF-IDF) method to construct a weighted coverage map. This map is key in calculating the fuzzing reward, which is then applied to the LLM-based mutator through reinforcement learning. CovRL-Fuzz, through this approach, enables the generation of test cases that are more likely to discover new coverage areas, thus improving vulnerability detection while minimizing syntax and semantic errors, all without needing extra post-processing.
Our evaluation results indicate that CovRL-Fuzz outperforms the state-of-the-art fuzzers in terms of code coverage and bug-finding capabilities: CovRL-Fuzz identified 48 real-world security-related bugs in the latest JavaScript engines, including 39 previously unknown vulnerabilities and 11 CVEs. 

\end{abstract}

\maketitle
\section{Introduction}
JavaScript (JS) engines are complex software components for parsing, interpreting, compiling, and executing JavaScript code in modern web browsers. These engines are essential for accessing today's interactive web and embedded applications. 
According to a recent survey, as of January 2024, JavaScript is employed as a client-side programming language by 98.9\% of web browsers~\cite{js_influence}. 
Given their extensive use and Turing-complete nature, securing \javascript{} engines is a critical requirement.
For instance, vulnerabilities in \javascript{} engines can lead to various attack patterns, encompassing threats such as information disclosure and the potential for bypassing web browser security measures~\cite{issue729991, shellonearth}.
Considering the high stakes, the need for continuous and automated testing, such as \textit{fuzzing}, is crucial for \javascript{} engines, despite challenges from their strict input grammar requirements.
Fuzzing involves providing invalid, unexpected, or random inputs, e.g., by mutation, to a program to detect bugs. 
Coverage-guided fuzzing, e.g., AFL~\cite{afl}, stands out as an effective method by using code coverage to guide the fuzzing process, ensuring a more thorough examination of the code paths and thereby increasing the chances of uncovering hidden bugs~\cite{afl}.

Previous research on fuzzing \javascript{} engines can broadly be divided into two main approaches: \textit{grammar-level} and \textit{token-level} fuzzing to deal with strict grammar. 
Grammar-level fuzzing techniques focus on producing inputs that are grammatically accurate~\cite{wang2017skyfire, IFuzzer, han2019codealchemist, patra2016learning, aschermann2019nautilus, die, sofi, wang2019superion}, while token-level fuzzing offers a more flexible method.
Token-level fuzzing transforms inputs into a sequence of tokens and then substitutes certain tokens without adhering strictly to grammar rules~\cite{salls2021token}. 
Coverage-guided fuzzing is also widely applied in fuzzing \javascript{} engines, encompassing both grammar-level and token-level fuzzing methods~\cite{wang2019superion, salls2021token, sofi, die}.

However, due to the continuous evolution of the JavaScript language, the grammar in \javascript{} engines is also being consistently updated to match these changes. Consequently, grammar-level fuzzing faces the challenge of needing to add new grammar rules frequently. Token-level fuzzing offers more flexibility compared to grammar-level fuzzing. Nevertheless, as mutations evolve from the initial seed, maintaining syntactical correctness becomes challenging, often leading to syntax errors. This limitation hinders the ability to uncover deeper bugs without inducing errors. Therefore, fuzzing \javascript{} engines requires mutating highly structured inputs, a task that traditional heuristic mutations alone find difficult in producing well-formed inputs. To overcome this, recent advancements have introduced fuzzing techniques that utilize Code-LLMs, capable of generating well-formed inputs (i.e., those that are syntactically informed) for compilers, deep learning libraries, and \javascript{} engines~\cite{ye2021comfort, titanfuzz, fuzzgpt, xia2023fuzz4all}.

Among these developments, TitanFuzz~\cite{titanfuzz} and FUZZ4ALL~\cite{xia2023fuzz4all} are notable for their use of Code-LLMs in mutation processes. Pretrained Code-LLMs, already trained on extensive datasets across various programming languages, can be directly employed for LLM-based mutation without the need for further finetuning.
Moreover, these models inherently understand the context of the language. This means they are capable of comprehending the grammar of the code and generating inputs that reflect both grammatical accuracy and contextual relevance. Their effectiveness is evident in their ability to generate seeds that are abundant in edge cases~\cite{ye2021comfort, fuzzgpt}.

While pretrained LLM-based mutators have proven to be effective for fuzzing~\cite{xia2023fuzz4all, titanfuzz}, it is important to note that all current LLM-based fuzzing approaches are categorized as black-box fuzzing and do not incorporate internal program information such as code coverage. 
In TitanFuzz~\cite{titanfuzz}, although a fitness function is used, it does not involve coverage-related information like in coverage-guided fuzzing. Instead, it utilizes static analysis information of the generated input, such as the number of unique function calls, depth, and iteration counts, as the fitness function's input. Therefore, TitanFuzz is not based on coverage-guided fuzzing, as it doesn't directly utilize execution code coverage information of the fuzzing target.

Differing from black-box fuzzing, coverage-guided fuzzing leverages internal program data. This method of fuzzing uses an evolutionary strategy to create interesting seeds that aim to enhance the target program's coverage. It considers the impact of mutated inputs on the program’s coverage. According to Miller's research, a 1\% increase in code coverage correlates to a 0.92\% higher probability of discovering bugs~\cite{fuzzbynumber}. Applying coverage data in coverage-guided fuzzing enhances coverage more effectively than black-box fuzzing, increasing the likelihood of discovering more bugs.
However, as we describe below, this is surprisingly challenging.

\noindent\textbf{Problem.}
When AFL's heuristic mutator is replaced with a pretrained LLM-based mutator in coverage-guided fuzzing, we observe a reduced error rate. 
However, this improvement did not translate into increased code coverage. 
Surprisingly, some performance patterns fell below that of random fuzzing.
This pattern was confirmed in our experiments, where we replaced AFL's mutator with an LLM-based mutator. Some of the LLM-based fuzzers obtained 12-16\% lower coverage in V8 and 4-13\% lower coverage in JerryScript compared to the baseline (random mutation). For more detailed results, please refer to Table~\ref{tab:ablation-study} in Section~\ref{subsection:ablation-study}.
Our results also align with findings from TitanFuzz's experiments. 
We speculate that this phenomenon arises because LLM-based mutators make constrained predictions. While AFL's interesting seeds are based on increased coverage, random mutators indiscriminately mutate tokens from the dictionary, regardless of context. In contrast, LLM-based mutators focus on contextually relevant token predictions, which reduces diversity (due to an overemphasis on context, they often predict common sentences, diminishing diversity). 
Thus, in coverage-guided fuzzing, LLM-based mutators' context-aware mutations reduce errors but also limit diversity, rendering them less effective than random fuzzing.

\begin{figure}[t]
\begin{center}
\includegraphics[width=\linewidth]{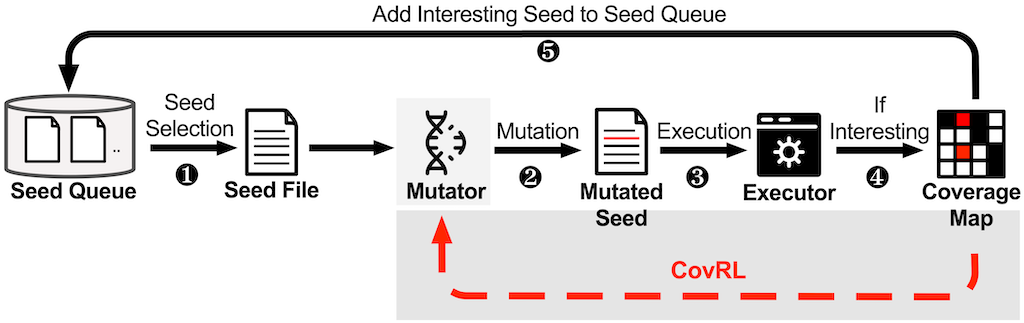}
\end{center}
\caption{\small{Overview: CovRL is a pioneering approach in integrating an LLM-based mutator into coverage-guided fuzzing.}}
\label{fig:cfg}
\end{figure}

\smallskip\noindent\textbf{Our Approach.} 
To address the aforementioned problem, we propose enhancing LLM-based mutators to align better with coverage-guided fuzzing. 
This involves a novel strategy of providing direct coverage feedback to the LLM-based mutator for more effective \javascript{} engine fuzzing as illustrated in Figure~\ref{fig:cfg}. 
Key to this approach is the use of a coverage-based weight map, where weights are assigned according to the inverse frequency of each coverage occurrence. 
By leveraging Term Frequency-Inverse Document Frequency (TF-IDF) for weighting the coverage map, the coverage-weighted reward is directly applied in reinforcement learning, enabling the LLM-based mutator to generate test cases that can achieve new coverage. 
This method enhances the model's ability to discover unknown vulnerabilities and reduces syntax and semantic errors without the need for additional post-processing.
We call this approach \textit{\CovRL{}-fuzz}, and unlike other LLM-based fuzzing techniques~\cite{ye2021comfort, titanfuzz, fuzzgpt, xia2023fuzz4all}, \CovRL{} is the first method that properly integrates an LLM-based mutator to coverage-guided fuzzing.

To sum up, this paper makes the following contributions:
\begin{itemize}[leftmargin=*]
\item We introduce \CovRL{}, a novel technique that combines LLMs with reinforcement learning from coverage feedback. This is a unique approach that directly integrates coverage feedback into the LLM using TF-IDF for advanced coverage-guided fuzzing.
\item We implement \CovRL{}-Fuzz, a coverage-guided fuzzer employing the \CovRL{} technique, focused on \javascript{} engine fuzzing. Our experiments show that \CovRL{}-Fuzz outperforms existing fuzzers in terms of code coverage and bug-finding capabilities. This advancement underscores CovRL-Fuzz's efficiency in navigating the complexities of \javascript{} engine fuzzing.
\item \CovRL{}-Fuzz successfully identified \bugTotal~real-world security-related bugs, including \bugUndiscover~previously unknown bugs (\bugCve~CVEs) in the latest \javascript{} engines.
\item To foster future research, we will open-source our work at publication time.
\end{itemize}

\section{Background and Related Work}

\subsection{\javascript{} Engine Fuzzing}
Fuzzing is a powerful automated bug-finding technique that generates and executes numerous inputs to identify vulnerabilities, crashes, or unexpected behaviors in software~\cite{miller1990empirical}. Both academia and industry have recognized its effectiveness in uncovering software bugs. However, fuzzing faces challenges with JS engines that require strict grammar for input. When the input is not syntactically correct, the JS engine returns a syntax error. On the other hand, semantic inconsistencies (e.g., errors with reference, type, range, or URI) lead to semantic errors~\cite{han2019codealchemist}. In both cases, the JS engine's core logic, which may contain hidden vulnerabilities, isn't executed.

To address these challenges, researchers have proposed grammar-level and token-level fuzzing approaches. These strategies employ heuristic methods to tackle the issue. The grammar-level technique transforms the seed into an Intermediate Representation (IR) to produce grammatically accurate inputs~\cite{wang2017skyfire, IFuzzer, han2019codealchemist, patra2016learning, aschermann2019nautilus, die, sofi, wang2019superion}. While this approach reliably produces syntactically correct inputs, it doesn't consistently account for semantic constraints. Moreover, it often demands substantial manual effort to craft the necessary grammar rules.
The token-level fuzzing approach~\cite{salls2021token} offers a more flexible method, free from the constraints of grammar rules. This technique transforms inputs into a sequence of tokens and substitutes certain ones. Although this method has demonstrated effectiveness in bug detection, its strategy of randomly replacing tokens—without accounting for inter-token relationships—places a significant dependency on the quality of the initial seed. Consequently, the approach often results in inputs that are not syntactically correct.

Recently, there has been a growing interest among researchers in utilizing deep learning-based Language Models (LM) in fuzzing, aiming to overcome the limitations of traditional fuzzing methods. Early endeavors leveraged RNN-based Language Models to mutate portions of inputs~\cite{lee2020montage, liu2019deepfuzz, godefroid2017learn, cummins2018deepsmith}. More recently, there's been a discernible trend towards the adoption of Large Language Models (LLMs) for seed generation and mutation~\cite{ye2021comfort, titanfuzz, fuzzgpt, xia2023fuzz4all}.
TitanFuzz~\cite{titanfuzz} is a black-box fuzzing technique to use LLMs for mutation, demonstrating the effectiveness of pretrained LLMs not only for seed generation but also for mutation.
While TitanFuzz didn't use internal target information like coverage in fuzzing, it did utilize static analysis metrics such as the number of unique function calls, depth, and iteration counts. These metrics, gathered from statically analyzing mutated inputs, helped select interesting seeds for fuzzing processes.
They substituted portions of inputs with a mask token for the mutation (we refer to ‘‘Mask Mutation'').
Conversely, FUZZ4ALL~\cite{xia2023fuzz4all} introduced mutation by adding mutation prompts, such as ‘\texttt{Please create a mutated program}' instead of using mask tokens.

\smallskip\noindent\textbf{Coverage-guided Fuzzing.} 
By leveraging coverage feedback to explore diverse code paths, coverage-guided fuzzing has consistently outperformed traditional black-box fuzzing in its ability to discover software bugs~\cite{fuzzbynumber}.
Tools like the American Fuzzy Lop (AFL) have notably shifted testing paradigms by focusing on maximizing code coverage, mutating, and generating input sequences. Such tools have been remarkably effective, uncovering a plethora of security issues and thereby validating their ability to detect vulnerabilities across a wide range of software~\cite{afl}. It has also been extensively applied in JS engine fuzzing, such as grammar-level and token-level fuzzing~\cite{wang2019superion, salls2021token, sofi, die}.

It is illustrated in Figure~\ref{fig:cfg}. Unlike black-box or white-box fuzzing, coverage-guided fuzzing utilizes code coverage information of the target software to explore varied code paths. This method initially requires the instrumentation of the software, leading to the creation of a coverage map—a matrix that tracks the frequency of specific code paths being accessed.

Following this setup, the fuzzing procedure begins with the selection of a seed from the seed queue for the mutation (\circled{1}). This selected seed is then mutated to generate a new test case (\circled{2}). Subsequently, the executor runs the target software with this test case, measuring its associated code coverage (\circled{3}). 
If coverage is not already recorded in the coverage map, it is deemed as new coverage.
Identifying such new coverage elevates the mutated test case to the status of an `interesting seed' (\circled{4}), which is then queued back into the seed pool (\circled{5}).
By continuously reintroducing such interesting seeds and encouraging the discovery of novel coverages through iterative mutations, coverage-guided fuzzing effectively generates a wide array of diverse test cases, each targeting the exploration of unexplored code areas of the target software.

\SetKwInput{KwInput}{Input}
\SetKwInput{KwOutput}{Output}

\SetKwFunction{Fg}{\small{\textbf{GetReward}}}
\SetKwFunction{FgetCov}{\small{\textbf{GetCov}}}
\SetKwFunction{Fm}{\small{\textbf{RewardModelingProcess}}}
\SetKwFunction{Fcount}{\small{\textbf{Count}}}
\SetKwProg{Fn}{Function}{:}{}

\definecolor{mycolor}{RGB}{31, 31, 200}

\renewcommand{\KwSty}[1]{\textnormal{\textcolor{mycolor}{\bfseries #1}}\unskip}
\begin{algorithm}[t]
\caption{Coverage-Rate Rewarding (CRR)}
\label{alg:coverage_rate_reward}
\KwInput{testcase $T$}
\KwOutput{reward $R_{cov}$}
$Total_{cov}$ : Total Coverage Map (Accumulated)\\

\bigskip

\Fn{\Fg{$T$}}{
    \uIf{\small{\textbf{JS\_Engine}}($T$) is $SyntaxError$}{
        \Return {-1.0} \\
    }
    \uElseIf{\small{\textbf{JS\_Engine}}($T$) is $SemanticError$}{
        \Return {-0.5} \\
    }
    \Else{
        /* \small{\textbf{JS\_Engine}}($T$) is Passed */ \\
        $T_{cov}$ = \FgetCov($Input$) \\
        $Total_{cov}$ += $T_{cov}$ $\oplus$ $Total_{cov}$ \\
        \Return \Fcount($T_{cov}$) / \Fcount($Total_{cov}$)
    }
}

\medskip

\Fn{\Fm{$T$}}{
    $R_{cov}$ = \Fg($T$) \\
     $Output$ $\gets$ $R_{cov}$ \\
    \Return $Output$
}
\end{algorithm}

\smallskip\noindent\textbf{RL-based Fuzzing.} 
Unlike coverage-guided fuzzing, RL-based fuzzing approaches~\cite{alphaprog, fuzzboost, drf} seek to improve performance not by utilizing coverage for seed selection, but by incorporating code coverage feedback into deep learning models, such as deep neural networks (DNNs) and recurrent neural networks (RNNs). They provide feedback using each code coverage as a reward, and for this purpose, they process the code coverage into a quantified reward, which is the ratio of the current coverage relative to the total cumulative coverage. We refer to this as Coverage-Rate Rewarding (CRR).

The details of the coverage-rate rewarding procedure are shown in Algorithm~\ref{alg:coverage_rate_reward}. In the case of a syntax error or semantic error in the JS engine, a fixed penalty is given (Lines 3-6). These penalty approach is also commonly seen in other RL methods targeting Code-LLM~\cite{le2022coderl, liu2023rltf, ppocoder}. When passed through the JS engine, a CRR is calculated (Line 7-11). The CRR is calculated as a ratio of the current coverage relative to the total cumulative coverage.
This approach has inherent challenge, notably its failure to differentiate between newly discovered and pre-existing coverage, which means it awards high scores for MANY COVERED, including those previously covered.
In other words, even if a test case does not find new coverage, it can still receive a high score if it covers a substantial amount of existing coverage.
In this work, we propose a new rewarding approach that addresses this problem.

\begin{figure*}[t]
\begin{center}
\includegraphics[width=\linewidth/100*85]{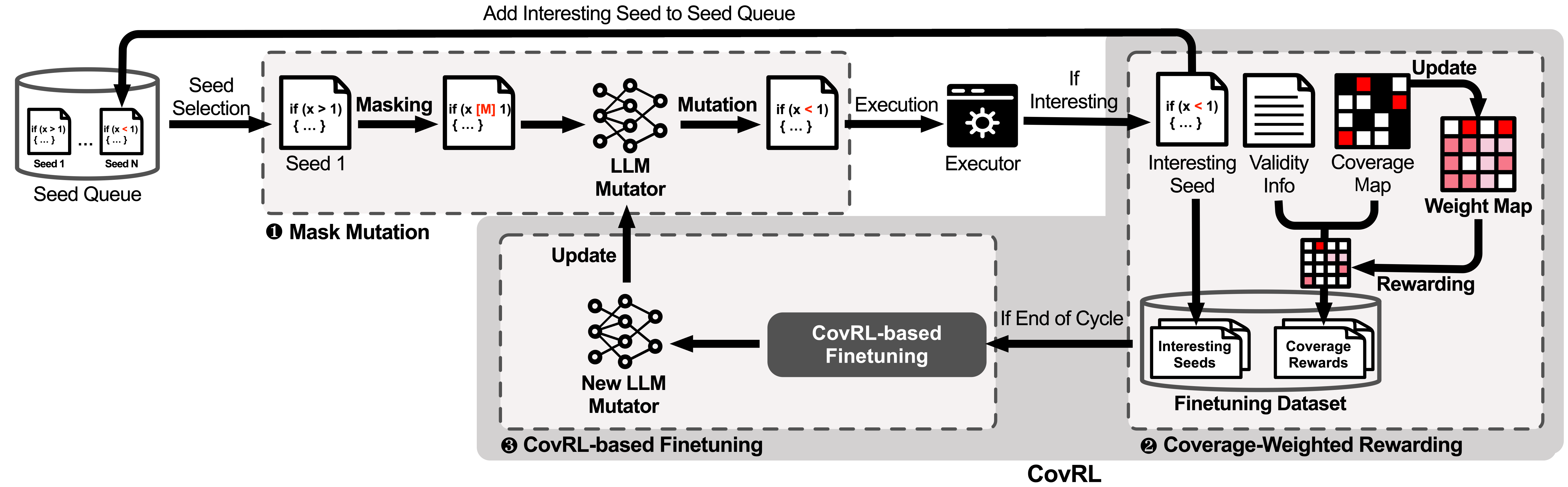}
\end{center}
\caption{\small{Workflow of \ourfuzzer{}: The gray-shaded area illustrates the operation of CovRL.}}
\label{fig:overview}
\end{figure*}

\subsection{Finetuning LLMs for code}

Following the success of LLMs in Natural Language Processing (NLP) tasks~\cite{2020gpt3, 2021reasoning, 2022palm}, the field of programming languages is advancing with significant contributions from Large Language Models for code (Code-LLMs) such as PLBART~\cite{ahmad2021plbart}, CodeT5~\cite{wang2021codet5, wang2023codet5+}, Codex~\cite{chen2021codex} and InCoder~\cite{fried2022incoder}. These advancements are facilitating various downstream tasks, including code completion~\cite{ziegler2022productivity}, program synthesis~\cite{le2022coderl, ppocoder, liu2023rltf, austin2021synthesis}, program repair~\cite{xia2022less, fan2023automated}, and many others.

Methods for finetuning LLMs, including Code-LLMs are categorized into supervised finetuning (SFT), instruction finetuning~\cite{wei2021FLAN}, and RL-based finetuning~\cite{rlhf, lee2023rlaif, roit2023rlef}. While prompt engineering controls the output of LLMs at inference time through input manipulation alone, SFT, instruction finetuning, and RL-based finetuning aim to steer the model during training time by learning from specific datasets tailored to particular tasks.
Particularly, 
RL-based finetuning has been proven effective in guiding LLMs using feedback to optimize factual consistency and reduce the toxic generation~\cite{rlhf, lee2023rlaif, roit2023rlef}. 
Recently, there has also been proposed for applying RL-based finetuning to Code-LLMs aimed at generating unit tests that are not only grammatically correct but also capable of solving complex coding tasks~\cite{le2022coderl, liu2023rltf, ppocoder}.

\smallskip\noindent\textbf{RL-based finetuning.}
RL-based finetuning consists of the following phases: reward modeling, and reinforcement learning. 
In reward modeling, an LLM-based rewarder is trained to evaluate the suitability of output results when input is provided to the LLM created in the previous phase. 
There are various approaches to feedback depending on how the rewarder is trained: utilizing an oracle~\cite{le2022coderl, ppocoder, liu2023rltf}, using deep learning models~\cite{lee2023rlaif}, and using human feedback~\cite{rlhf}. We also adopt the strategy of employing the JS engine as a feedback oracle.
In reinforcement learning, training commonly employs Kullback-Leibler (KL) divergence-based optimization. This method is designed to optimize the balance between maximizing rewards and minimizing deviation from the initial training distribution.

\section{Design}
In this section, we describe the design of \ourfuzzer{}. The key accomplishment of \ourfuzzer{} is:
We ensure effective fuzzing by limiting irrelevant mutations through context-aware mutations utilizing an LLM-based mutator and guiding the mutator with \CovRL{} to obtain a wide range of coverage.

Figure~\ref{fig:overview} is a workflow of \ourfuzzer{}, which consists of three phases. First, \ourfuzzer{} selects a seed from the seed queue. The seed undergoes a mask mutation process where specific tokens are masked and subsequently predicted (\circled{1}). We use mask tokens and predict their replacements using a masked language model task~\cite{devlin2018bert, raffel2020exploring}.
After mutation, the test case is then executed by the target \javascript{} engine. If the test case discovers new coverage not seen before, it is considered an interesting seed and is placed back in the seed queue for further mutation.
At the same time, \ourfuzzer{} stores the coverage map measured by the test case and validity information, whether the test case led to syntax errors, semantic errors, or passed successfully. Our rewarding approach uses validity information to impose penalties on inputs that result in syntax or semantic errors. Following this, it produces a rewarding signals by multiplying the current coverage map with a coverage-based weight map (\circled{2}).
After completing a mutation cycle, we proceed to finetune the LLM-based mutator using \CovRL{} by utilizing the gathered interesting seeds and rewarding signals.
We define the notion of one cycle as a predetermined number of mutations. 
The \CovRL{} employs the PPO~\cite{schulmanppo} algorithm, a method that seeks to improve the current model while adhering closely to the previous model's framework. The signal during training prevents the LLM from making syntax or semantic errors and induces prediction to find new coverage (\circled{3}).

Note that we do not perform any heuristic post-processing on the LLM-based mutator, save for \CovRL{}-based finetuning. We demonstrated a minimal error rate in using solely \CovRL{} that is comparable to other latest \javascript{} engine fuzzing techniques on Section~\ref{sec:RQ1}.

\subsection{Phase 1. Mask Mutation}
\label{sec:phase1_mask_mutation}
We use the mask mutation as a basic type of LLM-based mutation that can be done without any further prompts.

\smallskip\noindent\textbf{Masking.} To mutate the selected seed, \ourfuzzer{} performs a masking strategy for mask mutation (\circled{1} in Figure~\ref{fig:overview}). Given the input sequence $W = \{w_1, w_2, .., w_n\}$, \ourfuzzer{} uses three masking techniques: insert, overwrite, and splice. The strategy results in the mask sequence $W^{MASK} = \{[MASK], w_3, .., w_k\}$ and the masked sequence $W^{\backslash MASK} = \{w_1, w_2, [MASK], .., w_n\}$. The detailed operations are described as follows:
\begin{description}[leftmargin=!, itemsep=0.2em, parsep=0.2em, labelwidth=\widthof{\bfseries Overwrite}]
\item[Insert]{Randomly select positions and insert [MASK] tokens into the inputs.}
\item[Overwrite]{Randomly select positions and replace existing tokens with the [MASK] token.}
\item[Splice]{Statements within a seed are randomly divided into segments. A portion of these segments is replaced with a segment from another seed with [MASK], formatted as [MASK] statement [MASK].}
\end{description}

\smallskip\noindent\textbf{Mutation.} 
After generating a masked sequence $W^{\backslash MASK}$ via masking, the input is mutated by inferring in the masked positions via LLM-based mutator. The mutation design of \ourfuzzer{} is based on a span-based masked language model (MLM) that can predict variable-length masks~\cite{raffel2020exploring, fried2022incoder}. Thus, the MLM loss we utilize for mutation can be represented as follows:

\begin{equation}
    L_{MLM}(\theta) = \sum_{i=1}^{k} -logP_{\theta} (w_i^{MASK}|w^{\backslash MASK}, w_{<i}^{MASK})
\label{eq:MLM}
\end{equation}

$\theta$ represents the model's trainable parameters that are optimized during the training process, and $k$ is the number of tokens in $W^{MASK}$. $w^{\backslash MASK}$ denotes the masked input tokens where certain tokens are replaced by mask tokens. $w^{MASK}$ refers to the original tokens that have been substituted with the mask tokens in the input sequence. 

\subsection{Phase 2. Coverage-Weighted Rewarding}

\CovRL{} designs rewarding signal called Coverage-Weighted Rewarding (CWR)  for guiding the mutator.
The signal is weighted using TF-IDF~\cite{sparck1972tfidf} to prioritize the discovery of new coverage (\circled{2} in Figure~\ref{fig:overview}). 
The TF-IDF known as the statistical term-specificity method, computes the importance of a word to a document in a corpus, accounting for the fact that specific words appear more frequently in general. It is often used as a weight vector in information retrieval and text mining searches. We utilize it in constructing the coverage-based weight map.

\smallskip\noindent\textbf{Rewarding.} Enhancing the concepts from previous RL-based finetuning methods using Code-LLM~\cite{le2022coderl, ppocoder, liu2023rltf}, we extend the idea of using software output to apply a rewarding signal. Notably, errors in the \javascript{} engine can be broadly grouped into syntax errors and semantic errors, which include reference, type, range, and URI errors. 
Given that $W^{*}$ is the concatenation of the masked sequence $W^{\backslash MASK}$ and the mask sequence $W^{MASK}$, the following returns can be deduced based on input to the target:
\begin{equation}
\label{eq:testing_signal}
r(W^{*}) = 
\begin{cases}
    -1.0 & \text{if } W^{*} \text{ is syntax error} \\
    -0.5 & \text{if } W^{*}
    \text{ is semantic error} \\
    +R_{cov}& \text{if } W^{*} \text{ is passed} \\
\end{cases}
\end{equation}

In order to assist the LLM-based mutator in discovering new coverage, we provide an additional rewarding signal alongside Eq.~\ref{eq:testing_signal}.
Our approach focuses on the frequency of each coverage by assigning specific weights instead of using the CRR commonly used in traditional RL-based fuzzing. The rewarding procedure involves adjusting the coverage map by utilizing the TF-IDF weight map, calculating the weighted sum for each coverage information, and normalizing it to get scores.

At first, we noted that the coverage map is similar to the Term Frequency (TF) in that it calculates the frequency at which a specific coverage location is reached. However, with a \javascript{} engine, certain codes in the test case can trigger the same code coverage multiple times. A typical example is when the input includes repetitions such as 'a=1; a=1;'. This can result in duplicate triggers for the same coverage area.
In such cases, the importance of repetitive coverage is reduced. It emphasizes the need to differentiate between different types of coverage rather than merely focusing on how often it occurs. Therefore, we define the term $TF^{cov}$ as a map of unique coverage:

\begin{equation}
TF^{cov} = {\text{unique coverage map}}
\label{eq:tf}
\end{equation}
We define the coverage-based weight map $IDF^{cov}$ using the coverage map of each seed as follows:
\begin{equation}
\label{eq:idf}
IDF^{cov} = {1 \over \sqrt{M}}log({N \over 1 + DF^{cov}})
\end{equation}

where $N$ denotes the total number of unique coverage obtained. $DF^{cov}$ denotes the number of seeds that have achieved the specific coverage location. The weight map $IDF^{cov}$ is obtained by taking the inverse of $DF^{cov}$, resulting in greater weights for less common coverage. The variable $M$ denotes the overall size of the coverage map, which we utilized as a scale factor to adjust the weight value.

The reward is acquired by taking the weighted sum of $TF^{cov}$ and $IDF^{cov}$ to create the weighted coverage map, which is then weighted to obtain as
\begin{equation}
\label{eq:tfidf}
R_{TFIDF} = log(\sum_{i=1}^M tf_{i, t} \cdot idf_{i, t-1})
\end{equation}
where $t$ represents the current cycle. $tf_{i, t}$ refers to an element in $TF^{cov}_{t}$, and $idf_{i, t-1}$ refers to an element in $IDF^{cov}_{t-1}$ at the previous time step before updating the weights. Afterward, we proceed to normalize the findings as:

\begin{equation}
R_{cov} = 
\begin{cases}
     \sigma(R_{TFIDF}) & \text{if } R_{TFIDF} > 0 \\
    +0.5 & \text{otherwise}
\end{cases}
\label{eq:cov_reward}
\end{equation}
where $\sigma$ is a sigmoid function used to map $R_{TFIDF}$ to a value between 0 and 1.
If there's no new coverage and no error, we set $R_{cov}$ to 0.5 when it's zero or less to give the minimum reward. 
The $R_{cov}$ is calculated only if the test case is free from any syntax or semantic problems. Our rewarding scheme incentivizes the LLM-base mutator to explore a wider range of coverage by providing high payouts for test cases that achieve uncommon levels of coverage.

\smallskip\noindent\textbf{Update Weight Map with Momentum.} 
Following each cycle, \CovRL{} updates the IDF weight map. To mitigate dramatic changes in reward distribution, we use momentum at a rate of $\alpha$ to incorporate the prior weight while recalculating the map. The updated weight map is as follows:

\begin{equation}
\label{eq:weight_map}
IDF^{cov}_{t} = \alpha IDF^{cov}_{t-1} + (1- \alpha) IDF^{cov}_{t}
\end{equation}
where $IDF^{cov}_{t}$ means new weight map and $IDF^{cov}_{t-1}$ means previous weight map.

\SetKwInput{KwInput}{Input}
\SetKwInput{KwOutput}{Output}

\SetKwFunction{Fg}{\small{\textbf{GetReward}}}
\SetKwFunction{FgetCov}{\small{\textbf{GetCov}}}
\SetKwFunction{Fc}{\small{\textbf{CalcCovReward}}}
\SetKwFunction{Fup}{\small{\textbf{CalcIDF}}}
\SetKwFunction{FcalR}{\small{\textbf{CalcTFIDF}}}
\SetKwFunction{Fu}{\small{\textbf{GetUniqueCov}}}
\SetKwFunction{Fm}{\small{\textbf{RewardingProcess}}}
\SetKwProg{Fn}{Function}{:}{}

\definecolor{mycolor}{RGB}{31, 31, 200}

\renewcommand{\KwSty}[1]{\textnormal{\textcolor{mycolor}{\bfseries #1}}\unskip}
\begin{algorithm}[t]
\caption{Coverage-Weighted Rewarding (CWR)}
\label{alg:weighted_coverage_reward}
\KwInput{test case $T$}
\KwOutput{reward $R_{cov}$}
$TF_{t}^{cov}$ : Unique Coverage Map\\
$IDF_{t-1}^{cov}$ : Previous Weight Map\\
$IDF_{t}^{cov}$ : Weight Map \\
\bigskip

\Fn{\Fg{$T$}}{
    \uIf{\small{\textbf{JS\_Engine}}($T$) is $SyntaxError$}{
        \Return {-1.0} \\
    }
    \uElseIf{\small{\textbf{JS\_Engine}}($T$) is $SemanticError$}{
        \Return {-0.5} \\
    }
    \Else{
        /* \small{\textbf{JS\_Engine}}($T$) is Passed */ \\
        $T_{cov}$ = \FgetCov($T$) \\
        \Return \Fc($T_{cov}$) 
    }
}
\medskip

\Fn{\Fc{$T_{cov}$}}{
    $TF_{t}^{cov}$= \Fu($T_{cov}$) \\
    $IDF_{t}^{cov}$ $\gets$ \Fup($T_{cov}$) \\
    
    $R_{cov}$ $\gets$ \FcalR($TF_{t}^{cov}$, $IDF_{t-1}^{cov}$) \\   
    $IDF_{t-1}^{cov}$ $\gets$ $\alpha$ * $IDF_{t-1}^{cov}$ + (1 - $\alpha$) * $IDF_{t}^{cov}$ \\
    
    \medskip
    
    \eIf{$R_{cov}$ > 0}{
        \Return $R_{cov}$
    }{
        \Return $0.5$
    }
}

\medskip

\Fn{\Fm{$T$}}{
    $R_{cov}$ = \Fg($T$) \\
    $Output$ $\gets$ $R_{cov}$ \\
    \Return $Output$
}
\end{algorithm}

\smallskip\noindent\textbf{Coverage-Weighted Rewarding Algorithm.}
Algorithm~\ref{alg:weighted_coverage_reward} describes the overall procedure of CWR. The input is the mutated test case $T$, and the output is the reward $R_{cov}$ measured from $T$. Our objective is to compute the reward $R_{cov}$ by assigning weights to each coverage using the TF-IDF approach. Thus, we adhere to the subsequent course of action:

First, to evaluate the reward for the mutated test case $T$, we assess for syntax errors, semantic errors, and whether the test case was successfully executed in the \javascript{} engine.  If an error occurs, we impose a predetermined penalty, as shown in equation Eq.~\ref{eq:testing_signal} (Lines 5-8). Imposing the predetermined penalty allows LLM-based mutator to focus on minimizing errors, which is consistent with the strategy used in previous studies that apply RL to Code-LLMs~\cite{le2022coderl, ppocoder, liu2023rltf}. 
When $T$ is passed through the JS engine, we measure the coverage $T_{cov}$ of $T$ and calculate the reward based on this coverage (Lines 9-12). The procedure for constructing CWR is based on the TF-IDF weight map (Lines 14-21). $TF_t^{cov}$ is created by generating the unique coverage map of $T_{cov}$ (Line 14). Additionally, it calculates the $IDF_t^{cov}$ using the value of $T_{cov}$ (Line 15).

Subsequently, the TF-IDF-based reward $R_{cov}$ is calculated by using the weight map $IDF^{cov}_{t-1}$ that was created in the previous cycle and $TF^{cov}_{t}$ (line 16). 
The purpose of applying the weight map from the previous cycle to measure the reward is to assign higher scores to the newly obtained rewards based on the coverage distribution achieved in the previous cycle. If the reward $R_{cov}$ is greater than 0, it is returned as is; otherwise, a fixed value is returned (Lines 18-21). To mitigate significant changes in the reward distribution, we stabilize the reward by using $IDF^{cov}_{t-1}$ using a momentum rate of $\alpha$ (Line 17).
We demonstrate the effect of CWR with momentum in Section~\ref{subsection:ablation-study}.

\SetKwInput{KwInput}{Input}
\SetKwInput{KwOutput}{Output}

\SetKwFunction{Fcf}{\small{\textbf{Finetune\CovRL}}}
\SetKwFunction{FRewarder}{\small{\textbf{FinetuneRewarder}}}
\SetKwFunction{FMutator}{\small{\textbf{FinetuneMutator}}}
\SetKwFunction{FCMutator}{\small{\textbf{ChangeMutator}}}
\SetKwFunction{FCinteresting}{\small{\textbf{IsInteresting}}}
\SetKwFunction{FMaskMutation}{\small{\textbf{MaskMutation}}}
\SetKwFunction{FSelectSeed}{\small{\textbf{SelectSeed}}}
\SetKwFunction{Fuzzing}{\small{\textbf{FuzzOne}}}
\SetKwFunction{Fcwr}{\small{\textbf{RewardingProcess}}}
\SetKwProg{Fn}{Function}{:}{}

\definecolor{mycolor}{RGB}{31, 31, 200}

\renewcommand{\KwSty}[1]{\textnormal{\textcolor{mycolor}{\bfseries #1}}\unskip}
\begin{algorithm}[t]
\caption{Fuzzing with \CovRL}
\label{alg:rlcf}
\KwInput{finetuning dataset $D_{T}$}
$\mathcal{R}_{prev}$ : Previous LLM-based rewarder \\ 
$\mathcal{R}_{cur}$ : Current LLM-based rewarder \\ 
$\mathcal{M}_{prev}$ : Previous LLM-based mutator \\ 
$\mathcal{M}_{cur}$ : Current LLM-based mutator \\ 

\bigskip

\Fn{\Fuzzing{$seed\_queue$}}{
    \For{$i=1$ \KwTo $iter\_cycle$}{
        $seed$ $\gets$ \FSelectSeed{$seed\_queue$} \\
        $T$ $\gets$ \FMaskMutation{$\mathcal{M}_{cur}$, $seed$} \\
            
        \If{\FCinteresting{$T$}}{
            $T_{interest}$ $\gets$ $T$ \\
            $seed\_queue$.append($T_{interest}$) \\
            $R_{cov}$ = \Fcwr{$T_{interest}$} \\
            $data$ $\gets$ $T_{interest}$, $R_{cov}$ \\
                
            $D_T$.append($data$)
        }
    }

    \Fcf{$D_{T}$}
}
\medskip

\Fn{\Fcf{$D_{T}$}}{
    $\mathcal{R}_{prev}$, $\mathcal{M}_{prev}$ $\gets$ $\mathcal{R}_{cur}$, $\mathcal{M}_{cur}$ \\
    $\mathcal{R}_{cur}$ $\gets$ \FRewarder{$\mathcal{R}_{prev}$, $D_{T}$} \\
    $\mathcal{M}_{cur}$ $\gets$ \FMutator{$\mathcal{M}_{prev}$, $\mathcal{R}_{cur}$, $D_{T}$} \\
}

\end{algorithm}

\subsection{Phase 3. \CovRL{}-based Finetuning}
The fuzzing environment with mask mutation can be conceptualized as a bandit environment for RL. In this environment, a masked sequence $W^{\backslash MASK}$ is provided as input ($x$), and the expected output is a mask sequence $W^{MASK}$ ($y$).
Inspired by previous studies~\cite{le2022coderl, ppocoder, liu2023rltf}, we finetune our model using the PPO algorithm~\cite{schulmanppo}, an actor-critic reinforcement learning (\circled{3} in Figure~\ref{fig:overview}). In our situation, it can be implemented by finetuning two LLMs in tandem: one LLM acts as a mutator (actor), while the other LLM serves as a rewarder (critic).
We utilize a pretrained LLM to initialize the parameters both of mutator and rewarder. The rewarder is trained using the Eq.~\ref{eq:testing_signal}. It plays a crucial role in training the mutator.

For \CovRL{}-based finetuning with PPO, we define the \CovRL{} loss as following manner:
 \begin{equation}
\label{eq:rlcf1}
L_{\CovRL}(\theta) = - \mathbb{E}_{(x, y) \sim D_t}\left[R(x, y)_t \cdot \log\left(\frac{\pi_{\theta}^{t}(y|x)}{\pi^{t-1}(y|x)}\right) \right]
\end{equation}
where $R(x, y)$ represents the reward of \CovRL{}, and $D_t$ refers to the finetuning dataset that has been collected up to time step $t$. $\pi_{\theta}^{t}(y|x)$ with parameters $\theta$ is the trainable RL policy for the current mutator, and $\pi^{t-1}(y|x)$ represents the policy from the previous mutator.

To mitigate the overoptimization and maintain the LLM-based mutator's mask prediction ability, we also use KL regularization. The reward after adding the KL regularization is
\begin{equation}
R(x, y)_t = r(W^{*}) + \log\left(\frac{\pi_{\theta}^{t}(y|x)}{\pi^{t-1}(y|x)}\right)
\label{eq:rlcf2}
\end{equation}

\smallskip\noindent\textbf{Fuzzing with \CovRL{} Algorithm.}
Algorithm~\ref{alg:rlcf} details one cycle of the fuzzing loop with \CovRL{}. The cycle iterates for a predetermined number of $iter\_cycle$ (Lines 6-14).
The LLM-based mutator uses a seed chosen from the $seed\_queue$ to produce the test case $T$ (Lines 7-8). If $T$ is deemed a noteworthy seed, it is added to the seed queue and the reward for the particular $T_{interest}$ is calculated and added to $D_T$ (Lines 9-14). After completing these $iter\_cycle$ iterations, the gathered $D_{T}$ is utilized as training data to call the \texttt{\textbf{Finetune\CovRL{}}} function, which carries out \CovRL{}-based finetuning (Line 15). The procedure of \texttt{\textbf{Finetune\CovRL{}}} involves the finetuning of the LLM-based Rewarder $\mathcal{R}$ and the LLM-based Mutator $\mathcal{M}$ (Lines 17-19). Initially, we designate the existing model as $\mathcal{R}_{prev}$ and $\mathcal{M}_{prev}$ (Line 17).
Following that, $\mathcal{R}_{prev}$ is finetuned using the finetuning dataset $D_{T}$ to generate a new rewarder $\mathcal{R}_{cur}$ (Line 18). 
At this point, the rewarder has been trained to predict the rewarding signal as described in Eq.~\ref{eq:testing_signal}. By utilizing the finetuned $\mathcal{R}_{prev}$ and $D_{T}$, we finetune the mutator $\mathcal{M}_{prev}$ to generate $\mathcal{M}_{cur}$ (Line 19). For finetuning the mutator, we apply reward or penalty to the model using the \CovRL{} loss from Eq.~\ref{eq:rlcf1}.

\section{Implementation}
We implemented a prototype of \ourfuzzer{} using Pytorch v1.8 and afl 2.52b~\cite{afl}. 

\smallskip\noindent\textbf{Dataset.}
 We collected data from regression test suites in several repositories including V8, JavaScriptCore, ChakraCore, JerryScript, Test262~\cite{test262}, and js-vuln-db~\cite{js-vuln-db} as of December 2022. We then simply pre-processed the data for training data and seeds, resulting in a collection of 55,000 unique JavaScript files for our experiments. 

\smallskip\noindent\textbf{Pre-Processing.} 
We performed a simple pre-processing on the regression test suites of the JS engines mentioned above to remove comments, filter out grammatical errors, and simplify identifiers. We then used the processed data directly for training. The pre-processing was conducted utilizing the \texttt{-m} and \texttt{-b} options of the UglifyJS tool~\cite{uglifyjs}.

\smallskip\noindent\textbf{Training.}
We utilize the pretrained Code-LLM, CodeT5+ (220m)~\cite{wang2023codet5+}, as both the rewarder and the mutator. 
For the process of \CovRL{}-based finetuning, we trained the rewarder and mutator for 1 epoch each mutation cycle. We used a batch-size of 256 and learning rate of 1e-4. The optimization utilized the AdamW optimizer~\cite{loshchilov2017decoupled} together with a learning rate linear warmup technique.
Related experiments can be found in Table~\ref{tab:impact-epoch}.
The LLM-based rewarder uses the encoder from CodeT5+ to predict rewarding signal through a classification approach.    
we also employed the contrastive search technique~\cite{su2022contrastive}, applying a penalty factor $\alpha$ of 0.6 and setting the top-k of 32. The analysis of the optimal epoch and $\alpha$ selection in \CovRL{} can be found in Section~\ref{subsection:ablation-study}.
In addition, we align the coverage map size with AFL's recommendations by setting the scaling factor $M$ for the map size. This ensures that the instrumentational capacity is optimized. For moderate-sized software (approx. 10K lines), we employed a map size of $2^{16}$. For larger software exceeding 50K lines, we used a map size of $2^{17}$, striking a balance between granularity and performance. The number of lines in the target JS engine that we used can be located in Table~\ref{tab:lineofcode}.

\section{Evaluation}

To evaluate \ourfuzzer, we set three research questions.

\begin{itemize}[leftmargin=*]
\setlength\itemsep{0em}
\item\textbf{RQ1: } Is \ourfuzzer{} more effective and efficient than other state-of-the-art fuzzers?
\item \textbf{RQ2: } How does each component contribute to \ourfuzzer{}'s effectiveness?
\item \textbf{RQ3: } Can \ourfuzzer{} find real-world bugs in \javascript{} engines?
\end{itemize}

\subsection{Experimental Design}
\smallskip\noindent\textbf{Experimental setup.} Our setup included a 64-bit Ubuntu 20.04 LTS OS on an Intel(R) Xeon(R) Gold 6134 CPU @ 3.20GHz (64-core). Additionally, we harnessed three NVIDIA GeForce RTX 3090 GPUs for both training and mutation.

\begin{table}[t]
\centering
\caption{\small{Benchmarks of \javascript{} engines with their versions and lines of code}}
\resizebox{\columnwidth/100*72}{!}{
\tiny
\begin{tabular}{@{}lrr@{}}
\toprule
\textbf{JS Engine} & \textbf{Version}  & \textbf{\# of Lines} \\ 
\hline
V8                   & 11.4.73       & 1,087,873 \\
JavaScriptCore (JSC) & 2.38.1        & 566,262   \\
ChakraCore (Chakra)  & 1.13.0.0-beta & 782,996   \\
JerryScript (Jerry)  & 3.0.0         & 122,048   \\ \hline
QuickJS (QJS)             & 2021-03-27    & 75,257    \\
Jsish                & 3.5.0         & 58,143    \\
escargot             & bd95de3c, Jan 20 2024   & 311,473   \\
Espruino             & 2v20             & 26,945  \\
\bottomrule 
\end{tabular}
}
\label{tab:lineofcode}
\end{table}

\smallskip\noindent\textbf{Benchmarks.}
We tested it on four \javascript{} engines, using the latest versions as of January 2023: JavaScriptCore (2.38.1), ChakraCore (1.13.0.0-beta), V8 (11.4.73), JerryScript (3.0.0). 
We also conducted additional experiments on QuickJS, Jsish, escargot and Espruino for real bug detection experiments.
Table \ref{tab:lineofcode} presents the JS engine benchmarks used in our experiments. It displays both the version and the number of lines.

We built each target JS engine with Address Sanitizer (ASAN)~\cite{serebryany2012asan} to detect bugs related to abnormal memory access and with debug mode to find bugs related to undefined behavior.

\begin{table}[t]
\caption{\small{Baseline fuzzers targeting JS engines. CGF indicates the use of coverage-guided fuzzing, LLM denotes the usage of LLMs, and Mutation Level refers to the unit of mutation.}}
\label{tab:jsengine-studied-fuzzers}
\resizebox{\columnwidth/10*8}{!}{
\tiny
\begin{tabular}{@{}l|c|c|c|c@{}}

\toprule
\textbf{Fuzzer} & \textbf{CGF} & \textbf{LLM}  & \makecell{\textbf{Mutation} \\ \textbf{Level}} & \makecell{\textbf{Post} \\ \textbf{Processing}} \\ \midrule
\rowcolor{lightgray2}
\multicolumn{5}{l}{\textit{Heuristic Baselines}} \\
AFL(w/Dict)~\cite{afl} & \cmark &  & Bit/Byte &   \\
Superion~\cite{wang2019superion} & \cmark &  & Grammar & \\
Token-Level AFL~\cite{salls2021token} & \cmark &  & Token & \\
\rowcolor{lightgray2}
\multicolumn{5}{l}{\textit{LM Baselines}} \\
Montage~\cite{lee2020montage} &  &  & Grammar & \cmark\\
COMFORT~\cite{ye2021comfort} &  & \cmark & Grammar & \cmark \\ 
\hline
\textbf{\ourfuzzer} & \cmark & \cmark & Token &\\ \bottomrule 
\end{tabular}
}
\end{table}

\smallskip\noindent\textbf{Fuzzing Campaign.} For a fair evaluation, we used the same set of 100 valid seeds. For \textbf{RQ1}, we operated on 3 CPU cores considering other fuzzing approaches, and for \textbf{RQ2}, we used a single CPU core. 
For \textbf{RQ3}, we also used 3 CPU cores and conducted experiments for a week including four more JS engines apart from the four major ones.
To consider the randomness of fuzzing, we executed each fuzzer five times and then averaged the coverage results.
Also, to ensure fairness in fuzzing, the results of each experiment were measured, including the finetuning time through \CovRL{}. The average finetuning time is 10 minutes, occurring every 2.5 hours.

\smallskip\noindent\textbf{Baselines.} 
For \textbf{RQ1} and \textbf{RQ2}, we compare \ourfuzzer{} with state-of-the-art JS engine fuzzers, which include heuristic fuzzing techniques such as bit/byte-level fuzzing (AFL (w/Dict)~\cite{afl}), grammar-level fuzzing (superion~\cite{wang2019superion}), 
\begin{sloppy}
token-level fuzzing (Token-Level AFL~\cite{salls2021token}),
\end{sloppy}
and language model-based fuzzing techniques (Montage~\cite{lee2020montage}, COMFORT~\cite{ye2021comfort}).

In the case of Montage, it imports code from its test suite corpus, which might affect coverage by increasing the amount of executed code.
As a result, we included a version of Montage (w/o Import) in our experimental study, which does not import the other test suites. In the case of COMFORT, we evaluated it solely with the black-box fuzzer, excluding the differential testing component.
Each tool was run on four JS engines with default configurations which details can be found in Table~\ref{tab:jsengine-studied-fuzzers}.
Also, as part of our ablation study in \textbf{RQ2}, we performed an experiment comparing our approach to LLM-based fuzzers~\cite{titanfuzz, xia2023fuzz4all} that are not specifically designed for JS engines. This experiment involved using pretrained LLM-based mutation techniques, including mask mutation from TitanFuzz~\cite{titanfuzz} and prompt mutation from FUZZ4ALL~\cite{xia2023fuzz4all}.

\smallskip\noindent\textbf{Metrics.}
We use three metrics for evaluation.
\begin{itemize}[leftmargin=*]
    \item\textbf{Code Coverage} represents the range of the software's code that has been executed. We adopt edge coverage from the AFL's coverage bitmap, following FairFuzz~\cite{lemieux2018fairfuzz} and Evaluate-Fuzz-Testing~\cite{klees2018evaluating} settings. 
    We conducted a comparison of coverage in two categories: total and valid. Total refers to the coverage across all test cases, while valid refers to the coverage for valid test cases.
    We also employed the Mann-Whitney U-test~\cite{mann1947test} to assess the statistical significance and verified that all p-values were less than 0.05.
    \item\textbf{Error Rate} measures the rate of syntax errors and semantic errors in the generated test cases. This provides insight into how effectively each method explores the core logic in the target software. For detailed analysis, semantic errors are categorized into type errors, reference errors, URI errors, and internal errors based on the ECMA standard~\cite{ecma262}. 
    It should be noted that while COMFORT~\cite{ye2021comfort} utilized jshint~\cite{jshint} for measurement, focusing their error rate on syntax errors, we used JS engines, allowing us to measure the error rate including both syntax and semantic errors.
    \item\textbf{Bug Detection} is what the fuzzer is trying to find, which means a vulnerability. 
\end{itemize}

\subsection{RQ1. Comparison against existing fuzzers}
\label{sec:RQ1}

\begin{figure*}[t]
\centering
\vspace*{3.7cm}
\hspace{-5.5cm}
\begin{subfigure}{.24\textwidth}
    \centering
    \begin{tikzpicture}
    \includegraphics[scale=0.42]{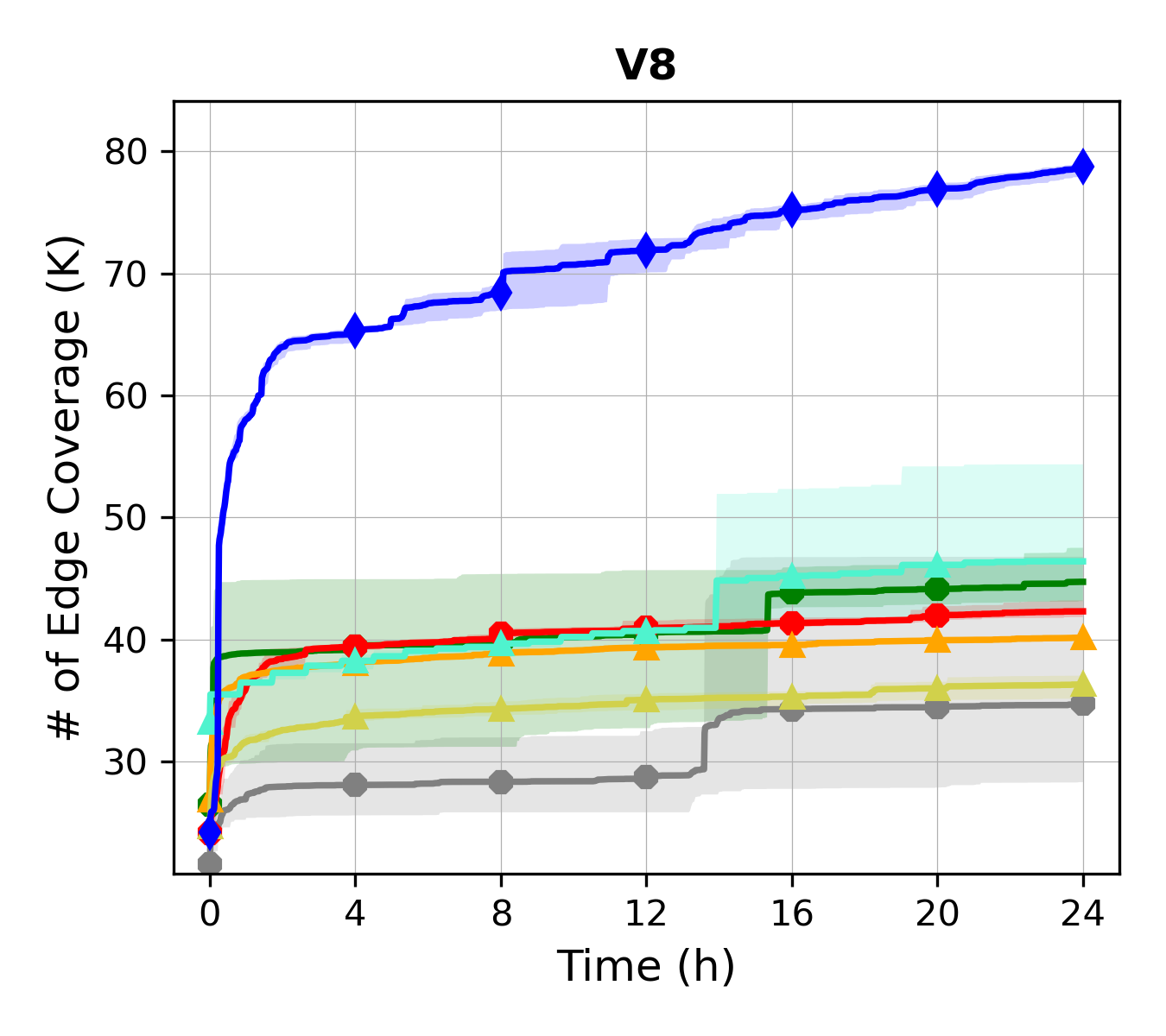}
    \end{tikzpicture}
\end{subfigure}
\hspace{0.2cm}
\begin{subfigure}{.24\textwidth}
    \centering
    \begin{tikzpicture}
    \includegraphics[scale=0.42]{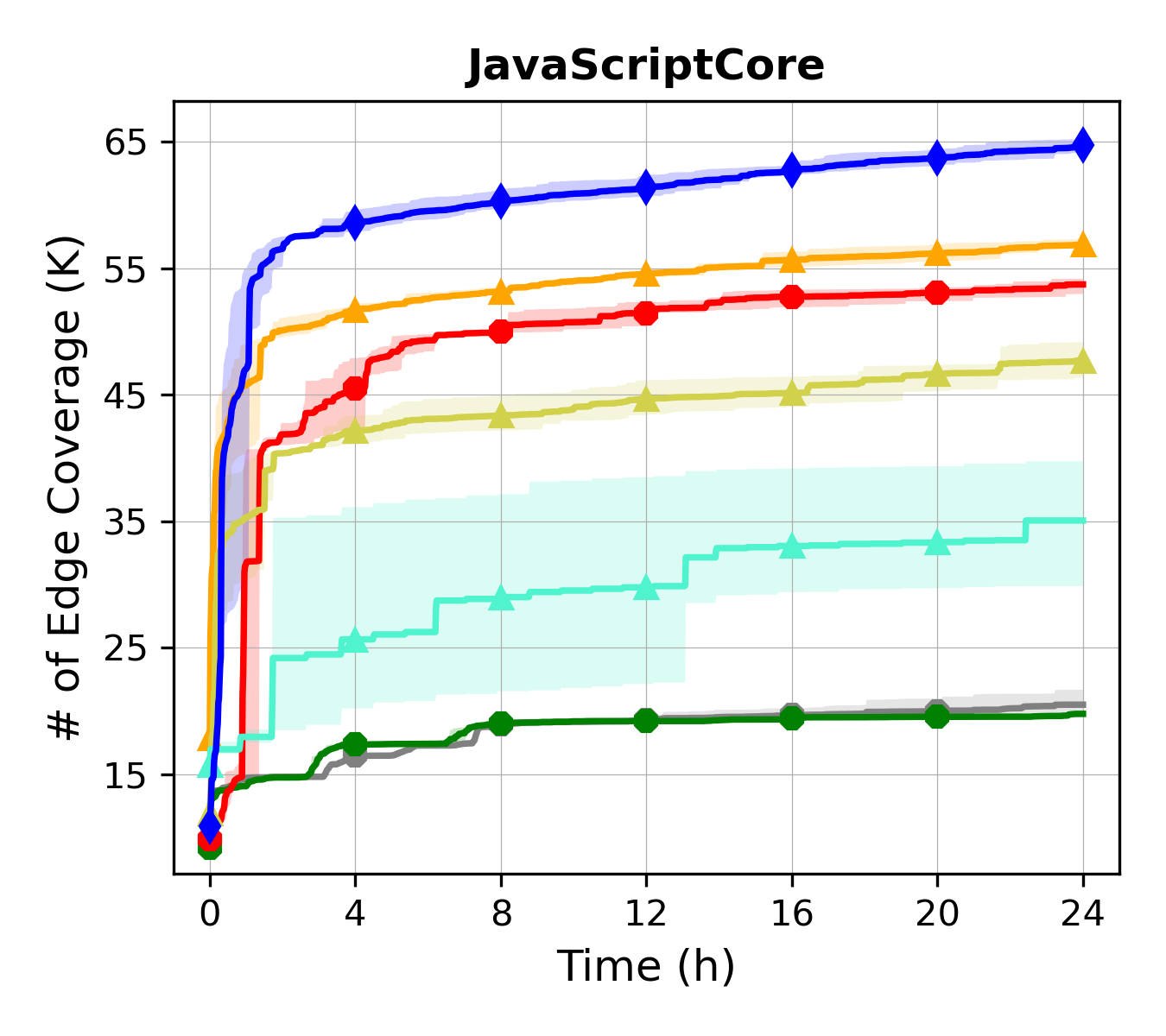}
    \end{tikzpicture}
\end{subfigure}
\hspace{0.2cm}
\begin{subfigure}{.24\textwidth}
    \centering
    \begin{tikzpicture}
    \includegraphics[scale=0.42]{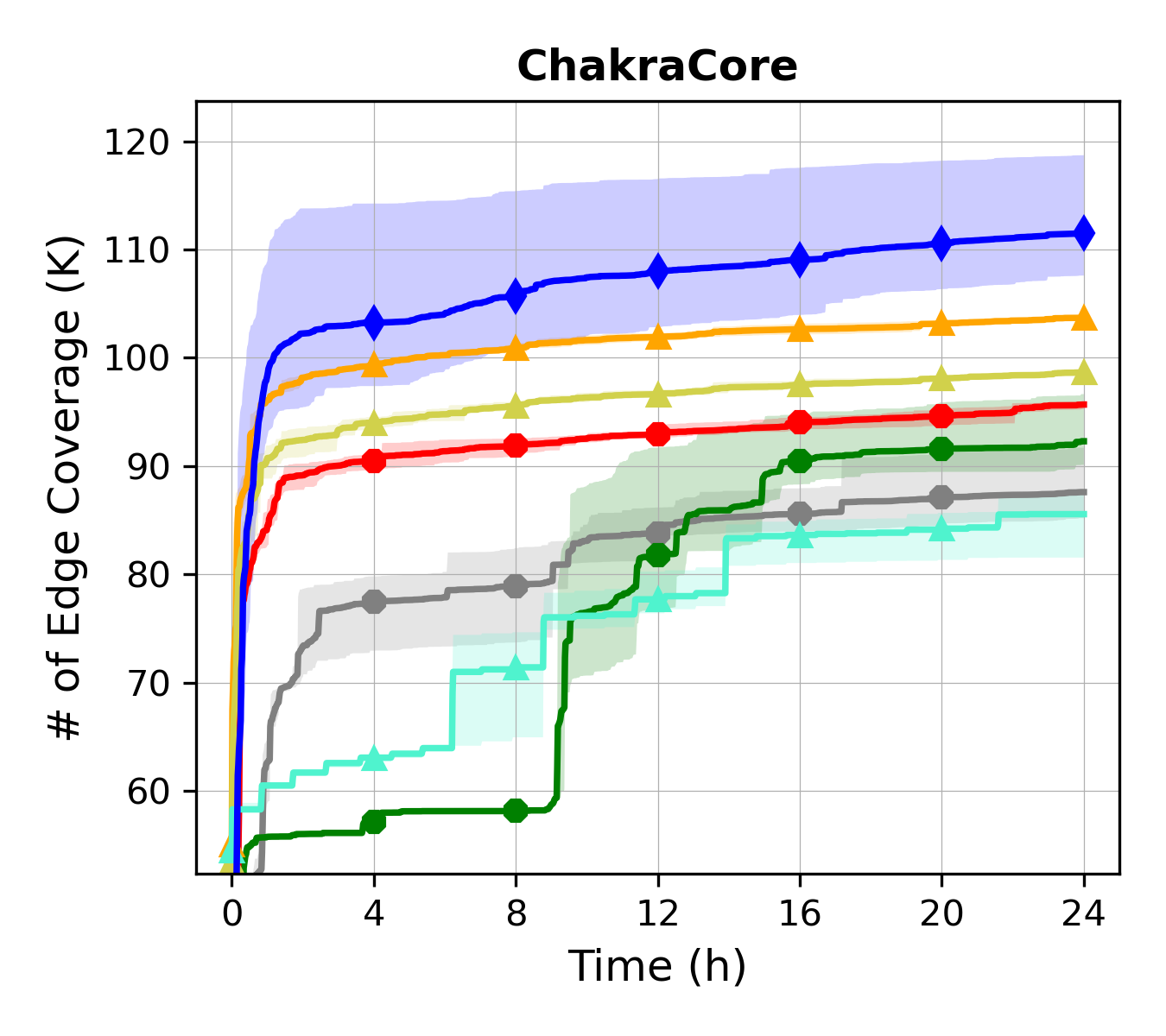}
    \end{tikzpicture}
\end{subfigure}
\hspace{0.2cm}
\begin{subfigure}{.24\textwidth}
    \centering
    \begin{tikzpicture}
    \includegraphics[scale=0.42]{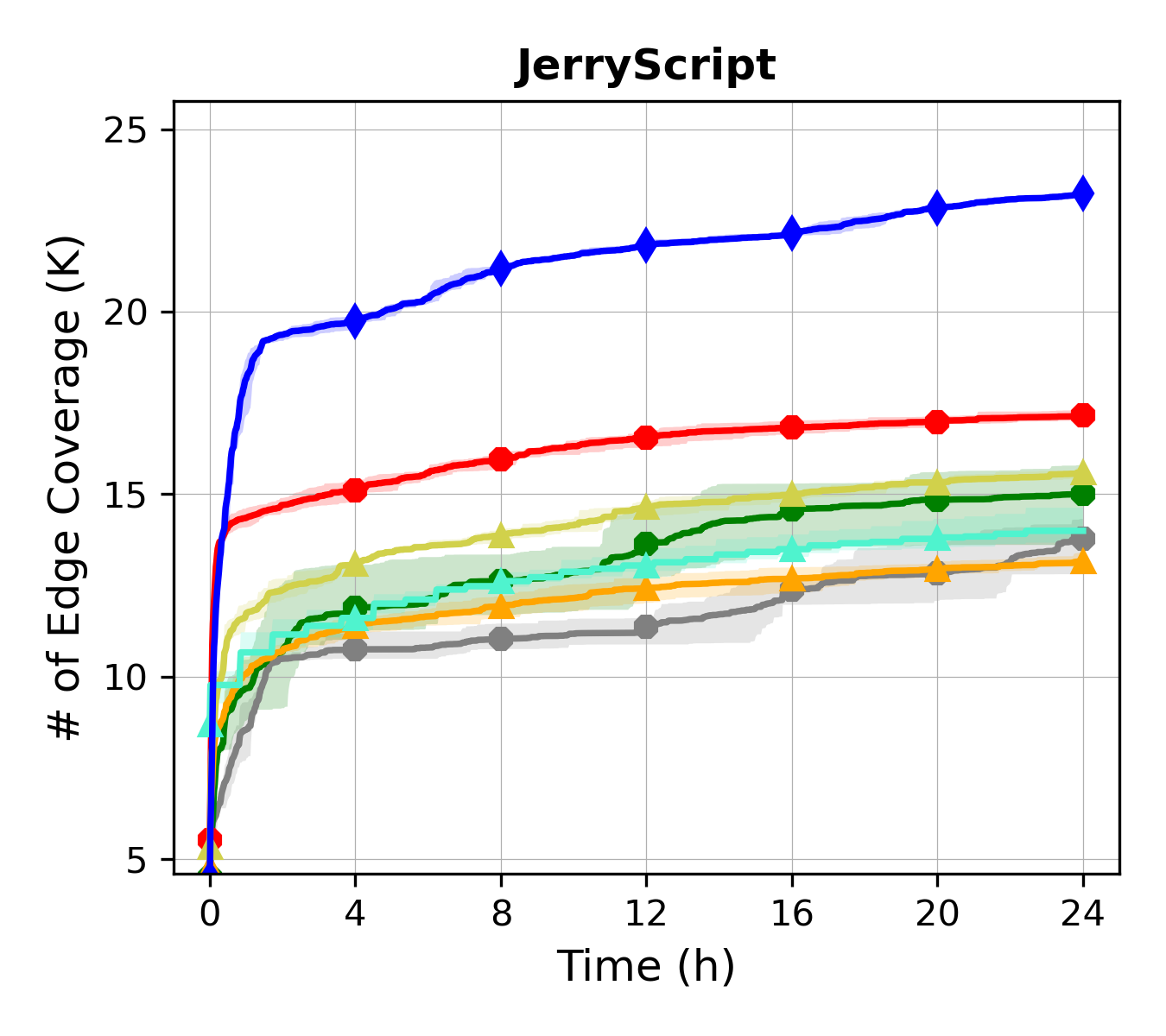}
    \end{tikzpicture}
\end{subfigure}

\hspace{-15cm}
\centering
\begin{tikzpicture}
\includegraphics[scale=0.45]{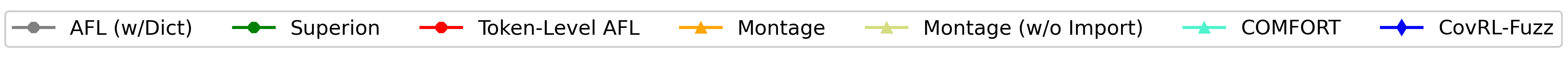}
\end{tikzpicture}
\vspace{-0.2cm}
\caption{\small{Number of edge coverage between \ourfuzzer{} and other \javascript{} engine fuzzers. The solid line represents the average coverage, while the shaded region depicts the range between the lowest and highest values five times.}}
\label{fig:24h_coverage}
\end{figure*}

\begin{table}[]
\caption{\small{Comparison with other \javascript{} engine fuzzers in Table~\ref{tab:jsengine-studied-fuzzers}. }}
\label{tab:coverage_quantative2}
\resizebox{\columnwidth}{!}{%
\begin{tabular}{@{}clrrrr@{}}
\hline \toprule 
\multirow{2}{*}{Target} & \multirow{2}{*}{Fuzzer} & \multirow{2}{*}{Error (\%)} & \multicolumn{2}{c}{Coverage} & \multicolumn{1}{c}{Improv} \\ 
    &   &   & \multicolumn{1}{c}{Valid} & \multicolumn{1}{c}{Total} &  \multicolumn{1}{c}{Ratio (\%)} \\ \hline
\multirow{6}{*}{V8}     
 & AFL (w/Dict)         & 96.90\%             & 29,929             & 33,531             & 134.79\% \\
 & Superion             & 77.35\%             & 33,812             & 36,985             & 112.87\% \\
 & Token-Level AFL      & 84.10\%             & 39,582             & 42,303             & 86.11\% \\
 & Montage              & \underline{56.24\%} & 38,856             & 40,155             & 96.06\% \\
 & Montage (w/o Import) & 94.08\%             & 33,487             &36,338              & 116.66\% \\
 & COMFORT              & 79.66\%             & \underline{44,324} & \underline{46,522} & 69.23\% \\
 & \ourfuzzer           & \textbf{48.68\%}    & \textbf{75,240}    & \textbf{78,729}    & \textbf{-} \\ \midrule
\multirow{6}{*}{JSC}    
 & AFL (w/Dict)         & 74.42\%             & 18,343             & 20,496             & 215.86\% \\
 & Superion             & 72.02\%             & 17,619             & 19,772             & 227.42\% \\
 & Token-Level AFL      & 69.70\%             & 52,385             & 53,719             & 20.51\% \\
 & Montage              & \textbf{42.34\%}    & \underline{55,511} & \underline{56,861} & 13.85\% \\
 & Montage (w/o Import) & 93.72\%             & 43,861             & 47,754             & 35.57\% \\
 & COMFORT              & 79.64\%             & 36,074             & 36,542             & 77.16\% \\
 & \ourfuzzer           & \underline{48.59\%} & \textbf{61,137}    & \textbf{64,738}    & \textbf{-} \\ \midrule
\multirow{6}{*}{Chakra} 
 & AFL (w/Dict)         & 81.32\%             & 83,038              & 87,587              & 27.30\% \\
 & Superion             & \textbf{42.63\%}    & 92,314              & 94,237              & 18.32\% \\
 & Token-Level AFL      & 90.64\%             & 92,621              & 95,677              & 16.54\% \\
 & Montage              & 82.21\%             & \underline{101,470} & \underline{103,589} & 7.63\% \\
 & Montage (w/o Import) & 94.72\%             & 90,940              & 98,643              & 13.03\% \\
 & COMFORT              & 79.47\%             & 81,171              & 83,142              & 34.11\% \\
 & \ourfuzzer           & \underline{54.87\%} & \textbf{105,121}    & \textbf{111,498}    & \textbf{-} \\ \midrule
\multirow{6}{*}{Jerry} 
 & AFL (w/Dict)         & \underline{77.32\%} & 9,307               & 14,259              & 63.03\% \\
 & Superion             & 86.23\%             & 8,944               & 15,061              & 54.35\% \\
 & Token-Level AFL      & 80.52\%             & \underline{14,361}  & \underline{17,152}  & 35.53\% \\
 & Montage              & 95.55\%             & 13,114              & 13,285              & 74.98\% \\
 & Montage (w/o Import) & 95.34\%             & 12,662              & 15,598              & 49.03\% \\
 & COMFORT              & 79.83\%             & 12,268              & 14,026              & 65.74\% \\ 
 & \ourfuzzer           & \textbf{58.84\%}    & \textbf{20,844}     & \textbf{23,246}     & \textbf{-} \\
 \bottomrule \hline
\end{tabular}
}
\end{table}

\begin{figure}[t]
\begin{center}
\includegraphics[width=1.0\linewidth]{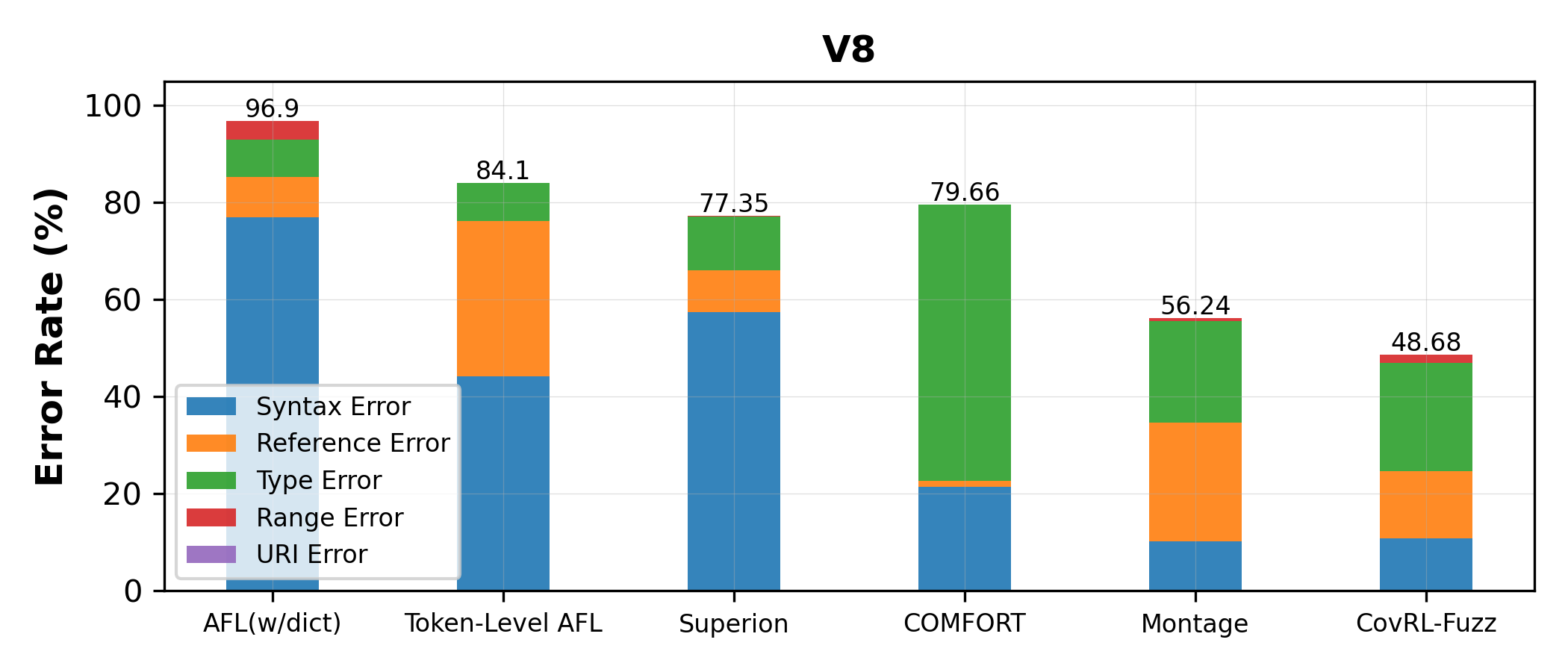}
\end{center}
\vspace*{-0.4cm}
\caption{\small{The error rate of generated test cases on V8. The four error types, excluding Syntax Error, are classified as Semantic Errors.}}
\label{fig:errorrate}
\vspace{-5mm}
\end{figure}
To answer \textbf{RQ1}, we ran all state-of-the-art heuristic and LM-based fuzzers listed in Table~\ref{tab:jsengine-studied-fuzzers} with the same 24-hour timeout, and we repeated the experiments five times to account for the randomness of fuzzing.

\smallskip\noindent\textbf{Code Coverage.}
Table~\ref{tab:coverage_quantative2} depicts the valid and total coverage for each fuzzing technique. The results of our evaluation demonstrate that \ourfuzzer{} outperforms state-of-the-art JS engine fuzzers. 
Our observation revealed that \ourfuzzer{} attained the highes coverage across all target engines, resulting in an average increase of 102.62\%/98.40\%/19.49\%/57.11\% in edge coverage.

To emphasize the effectiveness of \ourfuzzer{}, we monitored a growth trend of edge coverage, depicted in Figure~\ref{fig:24h_coverage}. In every experiment, \ourfuzzer{} consistently achieved the highest edge coverage more rapidly than any other fuzzer.
In contrast to heuristic baselines, \ourfuzzer{} immediately and significantly achieved higher coverage. This suggests that the LLM-based mutator of \ourfuzzer{} has a more potent ability to mutate than heuristic mutators for coverage-guided fuzzing.
\ourfuzzer{} also achieved high coverage compared to LM baselines. However, in ChakraCore, there was a marginal difference in coverage between Montage and \ourfuzzer{}, attributed to Montage’s strategy of importing and executing code from its test suite corpus, resulting in higher coverage. We observed that \ourfuzzer{} obtained significantly higher coverage when compared to Montage (w/o Import).

Note that, while other LM baselines did not account for training time, \ourfuzzer{} included the time required for \CovRL{} finetuning during the experiment. Additionally, we observed that \ourfuzzer{} continues to increase coverage when it nears the 24 hour mark. It displays its effectiveness in obtaining coverage.

\smallskip\noindent\textbf{Syntax and Semantic Correctness.}
\ourfuzzer{} is not a grammar-level fuzzing approach that prioritizes syntax and semantic validity.
However, it is assumed that \ourfuzzer{}, which uses reinforcement learning from a reward signal, can achieve higher validity compared to random fuzzing (such as Token-Level AFL).
To verify this assumption, we evaluate the error rate of unique test cases.

The experimental results are shown in Table~\ref{tab:coverage_quantative2}. \ourfuzzer{} demonstrated a lower error rate than Token-Level AFL for all JS engines. Furthermore, \ourfuzzer{} showed a lower error rate in comparison to most of the fuzzers. 
While it did not achieve the lowest error rate in JavaScriptCore and ChakraCore, \ourfuzzer{} still induced a significantly low error rate compared to the most of baselines.
Please note that the high error rate of Montage (w/o Import) is due to its inability to access functions from other test suites.

For a more detailed analysis of the error rate, we analyzed the types of errors triggered by fuzzers on V8, which is the most largest and dependable JS engine, as shown in Figure~\ref{fig:errorrate}.
The results showed that \ourfuzzer{} triggered fewer syntax errors in comparison to heuristic baselines. Furthermore, it also produced less syntax and semantic errors than LM baselines, even without using the post-processing techniques used by COMFORT and Montage.
These results indicate that \ourfuzzer{} is successful in reducing error rates exclusively through \CovRL{}, without requiring heuristic post-processing.

\smallskip\noindent\textbf{Finding bugs.} 
To study whether the coverage improvement and low error rate achieved by \ourfuzzer{} aid in detecting bugs, we conducted experiments by JS engines compiled in debug mode with ASAN. 
We relied on the output reports generated by ASAN for stack trace analysis to eliminate duplicate bugs.
We also manually analyzed and categorized these results by bug types.

Table~\ref{tab:finding_bugs_24h} shows the number and types of unique bugs found by the \ourfuzzer{} and compared fuzzers. \ourfuzzer{} discovered the most unique bugs compared to other fuzzers.
In detail, \ourfuzzer{} found 13 unique bugs and 8 of these bugs were exclusively detected by \ourfuzzer{}, including stack overflow and heap buffer overflow. 
These results highlight its effectiveness for bug detection.
As observed in the experimental results, LM-based fuzzers, despite achieving higher coverage, tend to find fewer bugs, while heuristic fuzzers, although achieving lower coverage, generally find more bugs. However, irrespective of this trend, \ourfuzzer{} demonstrated superior performance in effectively discovering the most bugs.

\begin{table}[]
\setlength{\tabcolsep}{2pt}
\caption{\small{Unique bugs discovered by \ourfuzzer{} and compared JS engine fuzzers.}}
\label{tab:finding_bugs_24h}
\resizebox{\columnwidth}{!}{
\begin{tabular}{@{}llcccccc@{}}
\hline
\toprule
\textbf{JS Engine} & \textbf{Bug Type} & \textbf{AFL} & \textbf{Superion} & \textbf{TokenAFL} & \textbf{Montage} & \textbf{COMFORT} & \textbf{\ourfuzzer} \\ \midrule
JSC & Undefined Behavior &  & \circledL{} &  &  &  & \circledL{} \\
JSC & Out-of-bounds Read &  &  &  &  &  & \circledL{} \\
Chakra & Undefined Behavior &  &  & \circledL{} &  & \circledL{} & \circledL{} \\
Chakra & Out of Memory &  &  &  &  &  & \circledL{} \\
Chakra & Out of Memory &  &  &  &  &  & \circledL{} \\
Jerry & Undefined Behavior & \circledL{} & \circledL{} & \circledL{} &  &  & \circledL{} \\
Jerry & Memory Leak & \small{} & \circledL{} & \circledL{} &  &  & \circledL{} \\
Jerry & Undefined Behavior & \circledL{} &  & \circledL{} &  &  & \circledL{} \\
Jerry & Undefined Behavior &  &  &  &  &  & \circledL{} \\ 
Jerry & Heap Buffer Overflow &  &  &  &  &  & \circledL{} \\
Jerry & Out of Memory &  &  &  &  &  & \circledL{} \\
Jerry & Stack Overflow &  &  &  &  &  & \circledL{} \\
Jerry & Undefined Behavior &  &  &  &  &  & \circledL{} \\
Jerry & Heap Buffer Overflow &  &  &  &  &  & \circledL{} \\ \midrule

SUM &  & 2 & 3 & 4 & 0 & 1 & 13 \\ \bottomrule \hline
\end{tabular}
}
\end{table}

\begin{table}[t]
\centering
\caption{\small{Variants of Ablation Study. Mutation Strategy refers to the method of mutation. Pretrained LLM denotes the Code-LLM used for mutation. Reward refers to the method of calculating rewards in \CovRL{}.}}
\resizebox{\columnwidth}{!}{%
\begin{tabular}{@{}l|c|c|c|c|c@{}}
\hline
\toprule
\textbf{Variants} & \textbf{CGF} & \textbf{Mutation Strategy} & \textbf{Pretrained LLM} & \textbf{\CovRL{}} & \textbf{Reward} \\ 
\midrule
Baseline (TokenAFL~\cite{salls2021token}) & \cmark & Random &  &   &  \\
\rowcolor{lightgray2}
\multicolumn{6}{l}{\textit{Pretrained LLM-based Mutators}} \\
Incoder w/Mask  & \cmark & Mask & Incoder (1B)~\cite{fried2022incoder} &  &  \\
StarCoder w/Mask  & \cmark & Mask & StarCoder (1B)~\cite{li2023starcoder} &  &  \\
StarCoder w/Prompt  & \cmark & Prompt & StarCoder (1B)~\cite{li2023starcoder} &  &  \\
CodeT5+ w/Mask & \cmark & Mask & CodeT5+ (220M)~\cite{wang2023codet5+} &  &  \\
\rowcolor{lightgray2}
\multicolumn{6}{l}{\textit{Finetuned LLM-based Mutators}} \\
SFT & \cmark & Mask & CodeT5+ (220M)~\cite{wang2023codet5+} &  &  \\
\CovRL{} w/CR & \cmark & Mask & CodeT5+ (220M)~\cite{wang2023codet5+} & \cmark & CR \\
\CovRL{} w/CRR & \cmark & Mask & CodeT5+ (220M)~\cite{wang2023codet5+} & \cmark & CRR \\
\midrule
\CovRL{}-Fuzz (w/CWR) & \cmark & Mask & CodeT5+ (220M)~\cite{wang2023codet5+} & \cmark & CWR \\ 
\bottomrule
\hline
\end{tabular}
}
\label{tab:var_ablation_study}
\end{table}

\subsection{RQ2. Ablation study}

\label{subsection:ablation-study}

\begin{table*}[t]
\caption{\small{The ablation study with each variant is detailed in Table~\ref{tab:var_ablation_study}. The Improv (\%) refers to the improvement ratio compared to the baseline.}}
\label{tab:ablation-study}
\resizebox{\textwidth}{!}{%
\begin{tabular}{l|crrr|crrr|crrr|crrr}
\hline
\toprule
\multicolumn{1}{c|}{Target} & \multicolumn{4}{c|}{V8} & \multicolumn{4}{c|}{JavaScriptCore} & \multicolumn{4}{c|}{ChakraCore} & \multicolumn{4}{c}{JerryScript} \\ \hline
\multicolumn{1}{c|}{\multirow{2}{*}{Variants}} & \multirow{2}{*}{Error (\%)} & \multicolumn{2}{c}{Coverage} & \multicolumn{1}{c|}{Improv} & \multirow{2}{*}{Error (\%)} & \multicolumn{2}{c}{Coverage} & \multicolumn{1}{c|}{Improv} & \multirow{2}{*}{Error (\%)} & \multicolumn{2}{c}{Coverage}  & \multicolumn{1}{c|}{Improv} & \multirow{2}{*}{Error (\%)} & \multicolumn{2}{c}{Coverage}  & \multicolumn{1}{c}{Improv} \\ 
    &   & \multicolumn{1}{c}{Valid} & \multicolumn{1}{c}{Total} & \multicolumn{1}{c|}{(\%)} &  & \multicolumn{1}{c}{Valid} & \multicolumn{1}{c}{Total} & \multicolumn{1}{c|}{(\%)} & & \multicolumn{1}{c}{Valid} & \multicolumn{1}{c}{Total} & \multicolumn{1}{c|}{(\%)} & & \multicolumn{1}{c}{Valid} & \multicolumn{1}{c}{Total} & \multicolumn{1}{c}{(\%)}\\ \hline
Baseline (TokenAFL~\cite{salls2021token}) & 88.79\% & 44,705 & 53,936 & - & 87.45\% & 35,406 & 37,461 & - & 78.98\% & 81,393 & 83,785 & - & 87.39\% & 12,312 & 14,795 & - \\
\rowcolor{lightgray2}
\multicolumn{17}{l}{\textit{Pretrained LLM-based Mutators}} \\
Incoder w/Mask & \textbf{49.08\%} & 46,427 & 47,385  & -12.15\% & \underline{50.31\%} & 44,191 & 44,643 & 19.17\% & \textbf{40.96\%} & 86,590 & 87,105 & 3.96\% & 77.63\% & 11,977 & 12,851 & -13.14\%\\
StarCoder w/Mask & 82.76\% & 53,779 & 56,256 & 4.30\% & 83.11\% & 42,007 & 43,136 & 15.15\% & 82.34\% & 86,988 & 88,842 & 6.04\% & 90.56\% & 12,174 & 13,817 & -6.61\%\\
StarCoder w/Prompt & 82.72\% & 41,331 & 45,034 & -16.50\% & 87.90\% & 45,545 & 47,568 & 26.98\% & 83.85\% & 84,597 & 87,351 & 4.26\% & 82.35\% & 13,777 & 15,413 & 4.18\%\\
CodeT5+ w/Mask & 62.68\% & 55,459 & 56,576 & 4.89\% & 55.40\% & 41,523 & 42,385 & 13.14\% & \underline{45.25\%} & 86,043 & 86,858 & 3.67\% & 78.48\% & 12,833 & 14,068 & -4.91\%\\
\rowcolor{lightgray2}
\multicolumn{17}{l}{\textit{Finetuned LLM-based Mutators}} \\
SFT & 74.91\% & \underline{58,230} & \underline{61,947} & 14.85\% & 65.57\% & 41,211 & 43,959 & 17.34\% & 69.96\% & 92,022 & \underline{95,334} & 13.78\% & 74.76\% & 15,927 & \underline{18,688} & 26.31\% \\
\CovRL{} w/CR & 71.77\% & 55,678 & 57,735 & 7.04\% & 53.00\% & \underline{47,116} & \underline{49,083} & 31.02\% & 67.50\% & \underline{92,465} & 94,145 & 12.36\% & \underline{73.35\%} & \underline{16,689} & 18,629 & 25.91\% \\
\CovRL{} w/CRR & 74.15\% & 57,401 & 61,331 & 13.71\% & 69.57\% & 37,230 & 43,369 & 15.77\% & 65.47\% & 91,427 & 94,785 & 13.13\% & 75.34\% & 16,118 & 18,584 & 25.61\%\\
\midrule
\ourfuzzer{} (w/CWR) & \underline{61.53\%} & \textbf{71,319} & \textbf{74,574} & \textbf{38.26\%} & \textbf{49.60\%} & \textbf{56,370} & \textbf{58,340} & \textbf{55.74\%} & 58.42\% & \textbf{96,257} & \textbf{98,221} & \textbf{17.23\%} & \textbf{58.59\%} & \textbf{17,481} & \textbf{19,855} & \textbf{34.20\%}\\
\bottomrule
\hline
\end{tabular}
}
\end{table*}

To answer \textbf{RQ2}, we conducted an ablation study on two key components of \ourfuzzer{}: (1) We compared pretrained LLM-based mutators and \ourfuzzer{} in terms of the type of LLMs and mutation strategies.
For this comparison, we utilized three LLMs and employed three mutation strategies.
Note that, as our focus is solely on LLM-based mutation, we did not include LLM-based generators in our study's scope, and thus they were not considered for comparison.

(2) Finetuned LLM-based mutators, we studied the impact of \CovRL{}-based finetuning on \ourfuzzer{}, focusing on the use of reward. The detailed configuration for the subject is as shown in Table~\ref{tab:var_ablation_study}. 
It is important to note that we only consider the impact of LLM-based mutators in the context of coverage-guided fuzzing. 
Therefore, all variations have been evaluated with only the mutation component being replaced, based on AFL.
Table~\ref{tab:ablation-study} shows the coverage and error rate of our studied variants, which were conducted by running them five times for five hours each, and the results were averaged.

\smallskip\noindent\textbf{Pretrained LLM-based Mutators.} 
We analyzed the results depending on the pretrained LLM and mutation strategies used for LLM-based mutators.
We studied CodeT5+ w/Mask, which is utilized in \ourfuzzer{}.
For comparison, we conducted a study using two other LLMs and mutation strategies: Incoder w/Mask, StarCoder w/Mask, and StarCoder w/Prompt. These two LLMs were used as pretrained LLM-based mutators in the TitanFuzz~\cite{titanfuzz} and Fuzz4ALL~\cite{xia2023fuzz4all} respectively, as control groups. We simply implemented the prompt mutation by adding mutation instructions (e.g. \texttt{Please create a mutated program that modifies the previous generation.}).

In the experimental results, we observed that the application of a pretrained LLM-based mutator, in comparison to the baseline which mutates randomly, resulted in a notable decrease in error rates. On the other hand, finetuned LLM-based mutators, including \ourfuzzer{}, consistently showed significantly higher coverage improvements compared to the baseline. This suggests that finetuning a pretrained LLM-based mutator is more effective for coverage-guided fuzzing than using it as is.
Additionally, we observed that the type and size of LLM did not have a substantial impact on the increase in coverage. Although the pretrained LLM of CodeT5+ w/Mask is just one-fifth the size of the other two models, the degree of improvement in coverage was not markedly different.

\smallskip\noindent\textbf{Finetuned LLM-based Mutators.} 
To demonstrate the effectiveness of \ourfuzzer{}, we froze the type and size of LLM and mutation strategy to examine the impact of different coverage rewards. The variants studied include: SFT, \CovRL{} w/CR, \CovRL{} w/CRR, and \ourfuzzer{} (w/CWR).
For SFT, we trained with our training dataset for the masked language model task. The experiment did not account for the training time, which was conducted independently. The training consisted of 10 epochs.

For rewarding, we additionally have designed a simple binary rewarding process, termed ``Coverage Reward (CR)''. Under this process, a reward of 1 is given to test cases that achieve new coverage, while a penalty of 0 is assigned to those that do not.

In the experimental results, \ourfuzzer{} achieved the highest coverage, both valid and total, compared to all control groups, and exhibited a low error rate. Furthermore, compared to baseline, \ourfuzzer{} showed an average increase of 36.36\% in total coverage and 44.75\% in valid coverage, while reducing the error rate by 28.62\%. This is a notable improvement, especially when compared to the other two rewarding processes, which were almost similar to SFT. Particularly considering that training time is not included for SFT, this demonstrates that \CovRL{}-Fuzz contributes not only to guiding the LLM to obtain more coverage but also to decreasing the error rate.

\begin{table}[t]
\caption{\small{Ablation : Impact for finetuning epochs}}
\resizebox{\columnwidth/2}{!}{
\label{tab:impact-epoch}
\begin{tabular}{@{}c|crr@{}}
\hline
\toprule 
 & \multicolumn{2}{c}{V8} \\ \midrule
\multirow{2}{*}{Epoch} & \multirow{2}{*}{Error (\%)} & \multicolumn{2}{c}{Coverage} \\ 
                 &                     & \multicolumn{1}{c}{Valid} & \multicolumn{1}{c}{Total} \\ \midrule
0 Epoch          & 74.91\%                   & 58,230              & 61,947 \\
\textbf{1 Epoch} & 61.53\%                   & \textbf{71,319}     & \textbf{74,574} \\
2 Epoch          & \underline{59.43\%}       & 66,017              & \underline{69,764} \\
3 Epoch          & \textbf{56.93\%}          & \underline{67,079}  & 69,517 \\ \bottomrule \hline
\end{tabular}
}
\end{table}

\begin{table}[t]
\caption{\small{Ablation : Impact for $\alpha$}}
\label{tab:impact-alpha}
\resizebox{0.9\columnwidth}{!}{
\begin{tabular}{@{}cc|c|c|c|c|c|c@{}}
\hline
\toprule 
\multicolumn{2}{c|}{$\alpha$} & 0.0 & 0.2 & 0.4 & \textbf{0.6} & 0.8 & 1.0 \\ \midrule
\multirow{2}{*}{Cov.}     & Valid & 68,248 & 68,635 & 69,247 & \textbf{71,319} & \underline{69,955} & 69,330 \\
                              & Total & 71,906 & 71,692 & 72,415 & \textbf{74,574} & \underline{72,623} & 72,218 \\
\bottomrule \hline
\end{tabular}
}
\end{table}

\smallskip\noindent\textbf{Impact Components.} 
We further conducted experiments on V8 to study two major impact components for \ourfuzzer{}: the \CovRL{}-based finetuning epochs and alpha. As with the earlier ablation studies, we conducted each experiment for 5 hours, repeated 5 times. In order to ensure fairness, any training time that exceeded one epoch was not included in the experiment duration for the \CovRL{}-based finetuning epochs.

Table~\ref{tab:impact-epoch} compares the coverage based on finetuning epochs. Our observation revealed a negative correlation between the number of epochs and the error rate, indicating that as the epochs rose, the error rate decreased. However, this decrease in error rate was also followed by a decrease in coverage. It indicates that overfitting starts at the second epoch, which may restrict the generation of diverse test cases.

Table~\ref{tab:impact-alpha} represents the comparison of coverage based on different values of $\alpha$. $\alpha$ refers to the momentum rate in Eq.~\ref{eq:weight_map}, which adjusts the weight between the previous and current $IDF^{cov}$. The experimental results demonstrated that applying a momentum rate of 0.6 led to better results compared to the absence of momentum.

\begin{lstfloat}
\begin{lstlisting}[language=JavaScript, label=lst:chakra_Out-of-Bounds-Read, caption=\small{The test case that triggers out-of-bounds read on ChakraCore 1.13.0.0-beta (\#13).}]
    function i ( t ) { }
    async function n ( t ) {
        if ( t instanceof i ) {
            let c = await i ( ) ;
            await c >> i ( n ) ;
        } else {
            var c = await n ( ) ;
        }
    }
    n ( true ) ;
\end{lstlisting}
\end{lstfloat}

\begin{lstfloat}
\begin{lstlisting}[language=JavaScript, label=lst:jerry_heap-buffer-overflow, caption=\small{The test case that triggers heap buffer overflow on JerryScript 3.0.0 (\#23).}]
    class s extends WeakMap { 
        static {} ; 
        } 
    function f ( )
\end{lstlisting}
\end{lstfloat}

\subsection{RQ3. Real-World Bugs}
In this section, we evaluated the ability of \ourfuzzer{} to find real-world bugs during a certain period of fuzzing. Specifically, we investigated how many real-world bugs \ourfuzzer{} can find and whether it can discover previously unknown vulnerabilities. 
Thus, we evaluated whether \ourfuzzer{} can find real-world bugs for 1 week for each target. We tested the latest version of each target engine as of January 2023 and found a total of \bugTotal{} bugs, including \bugUndiscover{} previously unknown vulnerabilities with \bugCve{} CVEs, some of which were internally fixed in the newer versions.

\begin{table*}[t]
\caption{\small{Summary of Detected Real-World Bugs: This table details \bugTotal{} bugs identified in JavaScript engines by our study (CovRL-Fuzz), including \bugCve{} that were classified as CVEs. Notably, \bugUndiscover{} of these bugs were previously unknown.
}}
\centering
\renewcommand{\arraystretch}{1.2}
\resizebox{\columnwidth*9/6}{!}{
\begin{tabular}{llllll}
\hline \toprule 

\textbf{\#}   & \textbf{JS Engine} & \textbf{Buggy Function} & \textbf{Bug Type}  & \textbf{Status}    & \textbf{Bug ID}     \\ \midrule
1   & V8         & NewFixedArray                  & Invalid size error       & Confirmed         & Issue * \\
2   & V8         & Builtins\_ArrayPrototypeSort   & Out of Memory     & Confirmed         & Issue * \\
\midrule
3   & JSC        & isSymbol                       & Out-of-bounds Read & Internal Fixed    & Bug *   \\
4   & JSC        & allocateBuffer                 & Crash by load()          & Confirmed         & Bug * \\
5   & JSC        & fixupArrayIndexOf              & Use After Free           & Internal Fixed    & Bug * \\
\midrule
6   & Chakra     & DeleteProperty                 & Undefined Behavior       & Confirmed         & Issue * \\
7   & Chakra     & PreVisitFunction               & Out of Memory            & Confirmed         & Issue * \\
8   & Chakra     & ParseDestructuredObjectLiteral & Undefined Behavior       & Reported          & Issue * \\
9   & Chakra     & RepeatCore                     & Out of Memory            & Reported          & Issue * \\
10  & Chakra     & GetSz                          & Out of Memory            & Reported          & Issue * \\
11  & Chakra     & UtcTimeFromStrCore             & Undefined Behavior       & Reported          & Issue * \\
12  & Chakra     & ToString                       & Undefined Behavior       & Reported          & Issue * \\
13  & Chakra     & TypePropertyCacheElement       & Out-of-bounds Read       & Reported          & Issue * \\
\midrule
14  & Jerry      & parser\_parse\_class           & Undefined Behavior       & Confirmed         & Issue * \\
15  & Jerry      & jmem\_heap\_finalize           & Undefined Behavior       & Confirmed         & Issue * \\
16  & Jerry      & parser\_parse\_statements      & Undefined Behavior       & Reported          & Issue * \\
17  & Jerry      & ecma\_builtin\_typedarray\_prototype\_sort   & Heap Buffer Overflow   & Reported    & CVE-*-* \\
18  & Jerry      & ecma\_regexp\_parse\_flags     & Undefined Behavior       & Reported          & CVE-*-* \\
19  & Jerry      & vm\_loop                       & Undefined Behavior       & Reported          & CVE-*-* \\
20  & Jerry      & ecma\_big\_uint\_div\_mod      & Undefined Behavior       & Reported          & CVE-*-* \\
21  & Jerry      & jmem\_heap\_alloc              & Out of Memory            & Reported          & CVE-*-* \\
22  & Jerry      & scanner\_literal\_is\_created  & Heap Buffer Overflow     & Reported          & CVE-*-* \\
23  & Jerry      & parser\_parse\_function\_statement   & Heap Buffer Overflow  & Reported       & CVE-*-* \\ 
24  & Jerry      & ecma\_property\_hashmap\_create      & Undefined Behavior    & Reported       & CVE-*-* \\
25  & Jerry      & parser\_parse\_for\_statement\_start & Undefined Behavior & Reported          & CVE-*-* \\
26  & Jerry      & jmem\_heap\_alloc              & Stack Overflow           & Reported          & Issue * \\
27  & Jerry      & scanner\_is\_context\_needed   & Heap Buffer Overflow     & Reported          & CVE-*-* \\
\midrule
28  & QJS        & js\_proxy\_isArray             & Stack Overflow           & Reported/Fixed             & CVE-*-* \\
\midrule
29  & Jsish      & jsiEvalCodeSub                 & Out-of-bounds Read       & Reported          & Issue * \\
30  & Jsish      & IterGetKeysCallback            & Stack Overflow           & Reported          & Issue * \\
31  & Jsish      & Jsi\_DecrRefCount              & Use After Free           & Reported          & Issue * \\
32  & Jsish      & SplitChar                      & Use After Free           & Reported          & Issue * \\
\midrule
33  & escargot   & parseLeftHandSideExpression    & Undefined Behavior       & Reported          & Issue *  \\
34  & escargot   & generateExpressionByteCode     & Undefined Behavior       & Reported          & Issue * \\
35  & escargot   & generateStatementByteCode      & Undefined Behavior       & Reported          & Issue * \\
36  & escargot   & hasRareData                    & Out-of-bounds Read       & Reported          & Issue * \\
37  & escargot   & readPointerIsNumberEncodedValue  & Out-of-bounds Read     & Reported          & Issue * \\
38  & escargot   & TightVector                    & Out-of-bounds Read       & Reported          & Issue * \\
39  & escargot   & setupAlternativeOffsets        & Stack Overflow           & Reported          & Issue * \\
40  & escargot   & setMutableBindingByBindingSlot & Undefined Behavior       & Reported          & Issue * \\
41  & escargot   & redefineOwnProperty            & Undefined Behavior       & Reported          & Issue * \\
42  & escargot   & asPointerValue                 & Undefined Behavior       & Reported          & Issue * \\
43  & escargot   & addOptionalChainingJumpPosition  & Undefined Behavior     & Reported          & Issue * \\
44  & escargot   & lastFoundPropertyIndex         & Stack Overflow           & Reported          & Issue * \\
45  & escargot   & setMutableBindingByIndex       & Undefined Behavior       & Reported          & Issue * \\
46  & escargot   & VectorCopier                   & memcpy-param-overlap     & Reported          & Issue * \\
\midrule
47  & Espruino   & jsvStringIteratorPrintfCallback   & Out-of-bounds Read    & Reported          & Issue * \\
48  & Espruino   & jspeFactorFunctionCall         & Stack Overflow           & Reported          & Issue *  \\

\bottomrule \hline
\end{tabular}
\label{tab:bug_finding}
}
\end{table*}

Table \ref{tab:bug_finding} illustrates a description of the discovered bugs. \textquotesingle Reported\textquotesingle\space in the Status column means that \ourfuzzer{} was the only fuzzer that discovered the bug, and it was reported because it persisted in the latest version. \textquotesingle Internal Fixed\textquotesingle\space refers to a bug that existed in a certain version but was not reported separately as a vulnerability and was fixed in the next version. If a bug was fixed after it was reported, it is labeled as \textquotesingle Reported/Fixed\textquotesingle. Additionally, if a bug was in the latest version despite being previously reported, it is labeled as \textquotesingle Confirmed\textquotesingle.
\ourfuzzer{} found a variety of bugs including undefined behaviors like assertion failures as well as memory vulnerabilities such as buffer overflow and use after free.
Note that, the experiment was carried out using only 3 cores and for a relatively short duration. In contrast, other fuzzing techniques have utilized an average of around 30 cores and have done their experiments for a whole month~\cite{lee2020montage, salls2021token, wang2019superion, ye2021comfort}. 

Despite these significant constraints, \ourfuzzer{} was still able to find a substantial number of unknown bugs. This suggests that \ourfuzzer{} demonstrated the effectiveness in finding real-world bugs on \javascript{} engines.

\smallskip\noindent\textbf{Case Study.}
Listing~\ref{lst:chakra_Out-of-Bounds-Read} represents a minimized test case generated by \ourfuzzer. This code triggered an out-of-bounds read bug in the ChakraCore 1.13.0, causing an abnormal termination of the \javascript{} engine. The original seed does not assign \texttt{await} to \texttt{var c}. \ourfuzzer{} changed it to \texttt{var c=await n();} and added the \texttt{await} statement on line 5, and also changed the condition of the \texttt{if} conditional. This caused the logic to call \texttt{await n();} repeatedly, which ultimately led to the bug.

Listing \ref{lst:jerry_heap-buffer-overflow} represents a minimized test case generated by \ourfuzzer, causing a heap buffer overflow in the release version of JerryScript 3.0.0. The bug occurs when a function declaration comes on the line following the declaration of a static initialization block in a class. 
When the parser read the statement, it didn't correctly distinguish the static initialization block range.
As a result, memory corruption occurred when parsing the function statement. In contrast to other fuzzing tools, \ourfuzzer{} is grammatically somewhat free and allows for context-aware mutation. this feature led to the discovery of this bug.
Our case study confirmed that these bugs can be only triggered by \ourfuzzer. 
This demonstrates the effectiveness of \ourfuzzer{} in detecting real-world bugs.

\section{Discussion}

We discuss three properties of \ourfuzzer{} in the following:

\smallskip\noindent\textbf{Diversity and validity.} 
To ensure diversity, we conducted experiments with seven fuzzers targeting four major \javascript{} engines such as V8, JavaScriptCore, ChakraCore, JerryScript.
Theoretically, adhering to syntax and semantics implies more constraints in mutation, which can make it more challenging to increase coverage. 
However, \ourfuzzer{} achieved higher coverage while maintaining low error rate of test cases (as shown in Figure~\ref{fig:24h_coverage} and Table~\ref{tab:coverage_quantative2}). 
It allowed that \ourfuzzer{} explore deeper code areas and detect more bugs compared to existing fuzzers.

\smallskip\noindent\textbf{Time spent between fuzzing and \CovRL{}-based finetuning.}
As mentioned in the experimental setup, we calculated the fuzzing time for the fairness of fuzzing, including the time spent on finetuning in the experimental results. 
On average, finetuning occurs for 10 minutes every 2.5 hours of fuzzing. 
Despite including the finetuning time in the experiment, \ourfuzzer{} achieved high coverage while also decreasing the error rate.

\smallskip\noindent\textbf{Supporting other targets.}
Through finetuning, the core idea of guiding coverage information directly with the LLM-based mutator is actually language-agnostic, which suggests its applicability to other language interpreters or compilers. 
However, our focus was more on analyzing the suitability of our idea to existing techniques than supporting various languages. 
Therefore, we conducted experiments only on \javascript{} engines, which we deemed to have the most impact. 
Extending to other targets is left as future work.

\section{Conclusion}
We introduced \ourfuzzer{}, a novel LLM-based coverage-guided fuzzing framework that integrates coverage-guided reinforcement learning for the first time. This approach enhances LLM-based fuzzing by leveraging coverage feedback to generate inputs that achieve broader coverage and deeper exploration of code logic without syntax limitations. Our evaluation results affirmed the superior efficacy of the \ourfuzzer{} methodology in comparison to existing fuzzing strategies. Impressively, it discovered \bugTotal{} real-world security-related bugs with \bugCve{} CVEs in JS engines — among these, \bugUndiscover{} were previously unknown vulnerabilities. We believe that our methodology paves the way for future studies focused on harnessing LLMs with coverage feedback for software testing.

\bibliographystyle{acm}
\bibliography{main}

\begin{thebibliography}{10}

\bibitem{ahmad2021plbart}
{\sc Ahmad, W., Chakraborty, S., Ray, B., and Chang, K.-W.}
\newblock Unified pre-training for program understanding and generation.
\newblock In {\em Proceedings of the 2021 Conference of the North American Chapter of the Association for Computational Linguistics: Human Language Technologies\/} (2021), pp.~2655--2668.

\bibitem{aschermann2019nautilus}
{\sc Aschermann, C., Frassetto, T., Holz, T., Jauernig, P., Sadeghi, A.-R., and Teuchert, D.}
\newblock Nautilus: Fishing for deep bugs with grammars.
\newblock In {\em Proceedings 2019 Network and Distributed System Security Symposium\/} (2019).

\bibitem{austin2021synthesis}
{\sc Austin, J., Odena, A., Nye, M., Bosma, M., Michalewski, H., Dohan, D., Jiang, E., Cai, C., Terry, M., Le, Q., et~al.}
\newblock Program synthesis with large language models, 2021.

\bibitem{drf}
{\sc B{\"o}ttinger, K., Godefroid, P., and Singh, R.}
\newblock Deep reinforcement fuzzing.
\newblock In {\em 2018 IEEE Security and Privacy Workshops (SPW)\/} (2018), IEEE, pp.~116--122.

\bibitem{2020gpt3}
{\sc Brown, T., Mann, B., Ryder, N., Subbiah, M., Kaplan, J.~D., Dhariwal, P., Neelakantan, A., Shyam, P., Sastry, G., Askell, A., et~al.}
\newblock Language models are few-shot learners.
\newblock In {\em Advances in neural information processing systems\/} (2020), vol.~33, pp.~1877--1901.

\bibitem{fuzzbynumber}
{\sc {Charlie Miller}}.
\newblock Fuzz by number.
\newblock \url{https://www.ise.io/wp-content/uploads/2019/11/cmiller_cansecwest2008.pdf}, 2008.
\newblock Accessed: 2024-01-12.

\bibitem{chen2021codex}
{\sc Chen, M., Tworek, J., Jun, H., Yuan, Q., Pinto, H. P. d.~O., Kaplan, J., Edwards, H., Burda, Y., Joseph, N., Brockman, G., et~al.}
\newblock Evaluating large language models trained on code, 2021.

\bibitem{2022palm}
{\sc Chowdhery, A., Narang, S., Devlin, J., Bosma, M., Mishra, G., Roberts, A., Barham, P., Chung, H.~W., Sutton, C., Gehrmann, S., et~al.}
\newblock Palm: Scaling language modeling with pathways, 2022.

\bibitem{2021reasoning}
{\sc Cobbe, K., Kosaraju, V., Bavarian, M., Chen, M., Jun, H., Kaiser, L., Plappert, M., Tworek, J., Hilton, J., Nakano, R., et~al.}
\newblock Training verifiers to solve math word problems, 2021.

\bibitem{cummins2018deepsmith}
{\sc Cummins, C., Petoumenos, P., Murray, A., and Leather, H.}
\newblock Compiler fuzzing through deep learning.
\newblock In {\em Proceedings of the 27th ACM SIGSOFT International Symposium on Software Testing and Analysis\/} (2018), pp.~95--105.

\bibitem{titanfuzz}
{\sc Deng, Y., Xia, C.~S., Peng, H., Yang, C., and Zhang, L.}
\newblock Large language models are zero-shot fuzzers: Fuzzing deep-learning libraries via large language models.
\newblock In {\em Proceedings of the 32nd ACM SIGSOFT International Symposium on Software Testing and Analysis (ISSTA 2023)\/} (2023).

\bibitem{fuzzgpt}
{\sc Deng, Y., Xia, C.~S., Yang, C., Zhang, S.~D., Yang, S., and Zhang, L.}
\newblock Large language models are edge-case fuzzers: Testing deep learning libraries via fuzzgpt, 2023.

\bibitem{devlin2018bert}
{\sc Devlin, J., Chang, M.-W., Lee, K., and Toutanova, K.}
\newblock Bert: Pre-training of deep bidirectional transformers for language understanding, 2018.

\bibitem{ecma262}
{\sc {ECMA International}}.
\newblock Ecmascript language speicification.
\newblock \url{https://www.ecma-international.org/ecma-262/}, 1997.
\newblock Accessed: 2023-08-15.

\bibitem{fan2023automated}
{\sc Fan, Z., Gao, X., Mirchev, M., Roychoudhury, A., and Tan, S.~H.}
\newblock Automated repair of programs from large language models.
\newblock In {\em 2023 IEEE/ACM 45th International Conference on Software Engineering (ICSE)\/} (2023), IEEE, pp.~1469--1481.

\bibitem{fried2022incoder}
{\sc Fried, D., Aghajanyan, A., Lin, J., Wang, S., Wallace, E., Shi, F., Zhong, R., Yih, S., Zettlemoyer, L., and Lewis, M.}
\newblock Incoder: A generative model for code infilling and synthesis.
\newblock In {\em The Eleventh International Conference on Learning Representations\/} (2022).

\bibitem{godefroid2017learn}
{\sc Godefroid, P., Peleg, H., and Singh, R.}
\newblock Learn\&fuzz: Machine learning for input fuzzing.
\newblock In {\em 2017 32nd IEEE/ACM International Conference on Automated Software Engineering (ASE)\/} (2017), ASE 2017, IEEE, pp.~50--59.

\bibitem{issue729991}
{\sc {Google}}.
\newblock Chrominum issue 729991.
\newblock \url{https://bugs.chromium.org/p/chromium/issues/detail?id=729991}, 2017.
\newblock Accessed: 2023-08-14.

\bibitem{han2019codealchemist}
{\sc Han, H., Oh, D., and Cha, S.~K.}
\newblock Codealchemist: Semantics-aware code generation to find vulnerabilities in javascript engines.
\newblock In {\em Proceedings 2019 Network and Distributed System Security Symposium\/} (2019).

\bibitem{sofi}
{\sc He, X., Xie, X., Li, Y., Sun, J., Li, F., Zou, W., Liu, Y., Yu, L., Zhou, J., Shi, W., et~al.}
\newblock Sofi: Reflection-augmented fuzzing for javascript engines.
\newblock In {\em Proceedings of the 2021 ACM SIGSAC Conference on Computer and Communications Security\/} (2021), ACM, pp.~2229--2242.

\bibitem{js-vuln-db}
{\sc {hoongwoo Han}}.
\newblock js-vuln-db.
\newblock \url{https://github.com/tunz/js-vuln-db}, 2010.
\newblock Accessed: 2023-08-15.

\bibitem{jshint}
{\sc {JSHint}}.
\newblock Jshint: A javascript code quality tool.
\newblock \url{https://jshint.com/}, 2013.
\newblock Accessed: 2023-08-15.

\bibitem{klees2018evaluating}
{\sc Klees, G., Ruef, A., Cooper, B., Wei, S., and Hicks, M.}
\newblock Evaluating fuzz testing.
\newblock In {\em Proceedings of the 2018 ACM SIGSAC conference on computer and communications security\/} (2018), ACM, pp.~2123--2138.

\bibitem{le2022coderl}
{\sc Le, H., Wang, Y., Gotmare, A.~D., Savarese, S., and Hoi, S. C.~H.}
\newblock Coderl: Mastering code generation through pretrained models and deep reinforcement learning.
\newblock {\em Advances in Neural Information Processing Systems 35\/} (2022), 21314--21328.

\bibitem{lee2023rlaif}
{\sc Lee, H., Phatale, S., Mansoor, H., Lu, K., Mesnard, T., Bishop, C., Carbune, V., and Rastogi, A.}
\newblock Rlaif: Scaling reinforcement learning from human feedback with ai feedback, 2023.

\bibitem{lee2020montage}
{\sc Lee, S., Han, H., Cha, S.~K., and Son, S.}
\newblock Montage: A neural network language $\{$Model-Guided$\}$$\{$JavaScript$\}$ engine fuzzer.
\newblock In {\em 29th USENIX Security Symposium (USENIX Security 20)\/} (2020), USENIX Association, pp.~2613--2630.

\bibitem{lemieux2018fairfuzz}
{\sc Lemieux, C., and Sen, K.}
\newblock Fairfuzz: A targeted mutation strategy for increasing greybox fuzz testing coverage.
\newblock In {\em Proceedings of the 33rd ACM/IEEE international conference on automated software engineering\/} (2018), pp.~475--485.

\bibitem{li2023starcoder}
{\sc Li, R., Allal, L.~B., Zi, Y., Muennighoff, N., Kocetkov, D., Mou, C., Marone, M., Akiki, C., Li, J., Chim, J., et~al.}
\newblock Starcoder: may the source be with you!
\newblock {\em arXiv preprint arXiv:2305.06161\/} (2023).

\bibitem{alphaprog}
{\sc Li, X., Liu, X., Chen, L., Prajapati, R., and Wu, D.}
\newblock Alphaprog: reinforcement generation of valid programs for compiler fuzzing.
\newblock In {\em Proceedings of the AAAI Conference on Artificial Intelligence\/} (2022), pp.~12559--12565.

\bibitem{fuzzboost}
{\sc Li, X., Liu, X., Chen, L., Prajapati, R., and Wu, D.}
\newblock Fuzzboost: Reinforcement compiler fuzzing.
\newblock In {\em Information and Communications Security: 24th International Conference, ICICS 2022, Canterbury, UK, September 5–8, 2022, Proceedings\/} (Berlin, Heidelberg, 2022), Springer-Verlag, p.~359–375.

\bibitem{liu2023rltf}
{\sc Liu, J., Zhu, Y., Xiao, K., Fu, Q., Han, X., Yang, W., and Ye, D.}
\newblock Rltf: Reinforcement learning from unit test feedback, 2023.

\bibitem{liu2019deepfuzz}
{\sc Liu, X., Li, X., Prajapati, R., and Wu, D.}
\newblock Deepfuzz: Automatic generation of syntax valid c programs for fuzz testing.
\newblock In {\em Proceedings of the AAAI Conference on Artificial Intelligence\/} (2019), pp.~1044--1051.

\bibitem{loshchilov2017decoupled}
{\sc Loshchilov, I., and Hutter, F.}
\newblock Decoupled weight decay regularization.
\newblock {\em arXiv preprint arXiv:1711.05101\/} (2017).

\bibitem{mann1947test}
{\sc Mann, H.~B., and Whitney, D.~R.}
\newblock On a test of whether one of two random variables is stochastically larger than the other.
\newblock {\em The annals of mathematical statistics\/} (1947), 50--60.

\bibitem{shellonearth}
{\sc Matt~Molinyawe, Adul-Aziz~Hariri, J.~S.}
\newblock \$hell on earth: From browser to system compromise.
\newblock In {\em Black Hat USA\/} (2016).

\bibitem{afl}
{\sc {Michal Zalewski}}.
\newblock Afl: American fuzzy lop.
\newblock \url{https://lcamtuf.coredump.cx/afl/}, 2013.
\newblock Accessed: 2023-08-15.

\bibitem{uglifyjs}
{\sc {Mihai Bazon}}.
\newblock uglifyjs.
\newblock \url{https://github.com/mishoo/UglifyJS}, 2010.
\newblock Accessed: 2023-08-14.

\bibitem{miller1990empirical}
{\sc Miller, B.~P., Fredriksen, L., and So, B.}
\newblock An empirical study of the reliability of unix utilities.
\newblock {\em Communications of the ACM 33}, 12 (Dec. 1990), 32--44.

\bibitem{rlhf}
{\sc Ouyang, L., Wu, J., Jiang, X., Almeida, D., Wainwright, C., Mishkin, P., Zhang, C., Agarwal, S., Slama, K., Ray, A., et~al.}
\newblock Training language models to follow instructions with human feedback.
\newblock {\em Advances in Neural Information Processing Systems 35\/} (2022), 27730--27744.

\bibitem{die}
{\sc Park, S., Xu, W., Yun, I., Jang, D., and Kim, T.}
\newblock Fuzzing javascript engines with aspect-preserving mutation.
\newblock In {\em 2020 IEEE Symposium on Security and Privacy (SP)\/} (2020), IEEE, pp.~1629--1642.

\bibitem{patra2016learning}
{\sc Patra, J., and Pradel, M.}
\newblock Learning to fuzz: Application-independent fuzz testing with probabilistic, generative models of input data.
\newblock Tech. rep., TU Darmstadt, Department of Computer Science, 2016.

\bibitem{raffel2020exploring}
{\sc Raffel, C., Shazeer, N., Roberts, A., Lee, K., Narang, S., Matena, M., Zhou, Y., Li, W., and Liu, P.~J.}
\newblock Exploring the limits of transfer learning with a unified text-to-text transformer.
\newblock {\em The Journal of Machine Learning Research 21}, 1 (Jan. 2020), 5485--5551.

\bibitem{roit2023rlef}
{\sc Roit, P., Ferret, J., Shani, L., Aharoni, R., Cideron, G., Dadashi, R., Geist, M., Girgin, S., Hussenot, L., Keller, O., et~al.}
\newblock Factually consistent summarization via reinforcement learning with textual entailment feedback, 2023.

\bibitem{salls2021token}
{\sc Salls, C., Jindal, C., Corina, J., Kruegel, C., and Vigna, G.}
\newblock $\{$Token-Level$\}$ fuzzing.
\newblock In {\em 30th USENIX Security Symposium (USENIX Security 21)\/} (2021), USENIX Association, pp.~2795--2809.

\bibitem{schulmanppo}
{\sc Schulman, J., Wolski, F., Dhariwal, P., Radford, A., and Klimov, O.}
\newblock Proximal policy optimization algorithms, 2017.

\bibitem{serebryany2012asan}
{\sc Serebryany, K., Bruening, D., Potapenko, A., and Vyukov, D.}
\newblock $\{$AddressSanitizer$\}$: A fast address sanity checker.
\newblock In {\em 2012 USENIX annual technical conference (USENIX ATC 12)\/} (2012), pp.~309--318.

\bibitem{ppocoder}
{\sc Shojaee, P., Jain, A., Tipirneni, S., and Reddy, C.~K.}
\newblock Execution-based code generation using deep reinforcement learning, 2023.

\bibitem{sparck1972tfidf}
{\sc Sparck~Jones, K.}
\newblock A statistical interpretation of term specificity and its application in retrieval.
\newblock {\em Journal of documentation 28}, 1 (1972), 11--21.

\bibitem{su2022contrastive}
{\sc Su, Y., Lan, T., Wang, Y., Yogatama, D., Kong, L., and Collier, N.}
\newblock A contrastive framework for neural text generation.
\newblock {\em Advances in Neural Information Processing Systems 35\/} (2022), 21548--21561.

\bibitem{test262}
{\sc {Technical Committee 39 ECMA International}}.
\newblock Test262.
\newblock \url{https://github.com/tc39/test262}, 2010.
\newblock Accessed: 2023-08-15.

\bibitem{IFuzzer}
{\sc Veggalam, S., Rawat, S., Haller, I., and Bos, H.}
\newblock Ifuzzer: An evolutionary interpreter fuzzer using genetic programming.
\newblock In {\em Computer Security--ESORICS 2016: 21st European Symposium on Research in Computer Security, Heraklion, Greece, September 26-30, 2016, Proceedings, Part I 21\/} (2016), Springer, Cham, pp.~581--601.

\bibitem{js_influence}
{\sc {W3Techs}}.
\newblock Usage statistics of javascript as client-side programming language on websites.
\newblock \url{https://w3techs.com/technologies/details/cp-javascript}, 2024.
\newblock Accessed: 2024-01-17.

\bibitem{wang2017skyfire}
{\sc Wang, J., Chen, B., Wei, L., and Liu, Y.}
\newblock Skyfire: Data-driven seed generation for fuzzing.
\newblock In {\em 2017 IEEE Symposium on Security and Privacy (SP)\/} (2017), IEEE, pp.~579--594.

\bibitem{wang2019superion}
{\sc Wang, J., Chen, B., Wei, L., and Liu, Y.}
\newblock Superion: Grammar-aware greybox fuzzing.
\newblock In {\em 2019 IEEE/ACM 41st International Conference on Software Engineering (ICSE)\/} (2019), IEEE, pp.~724--735.

\bibitem{wang2023codet5+}
{\sc Wang, Y., Le, H., Gotmare, A.~D., Bui, N.~D., Li, J., and Hoi, S.~C.}
\newblock Codet5+: Open code large language models for code understanding and generation, 2023.

\bibitem{wang2021codet5}
{\sc Wang, Y., Wang, W., Joty, S., and Hoi, S.~C.}
\newblock Codet5: Identifier-aware unified pre-trained encoder-decoder models for code understanding and generation.
\newblock In {\em Proceedings of the 2021 Conference on Empirical Methods in Natural Language Processing\/} (2021), pp.~8696--8708.

\bibitem{wei2021FLAN}
{\sc Wei, J., Bosma, M., Zhao, V., Guu, K., Yu, A.~W., Lester, B., Du, N., Dai, A.~M., and Le, Q.~V.}
\newblock Finetuned language models are zero-shot learners.
\newblock In {\em International Conference on Learning Representations\/} (2021).

\bibitem{xia2023fuzz4all}
{\sc Xia, C.~S., Paltenghi, M., Tian, J.~L., Pradel, M., and Zhang, L.}
\newblock Universal fuzzing via large language models.
\newblock {\em arXiv preprint arXiv:2308.04748\/} (2023).

\bibitem{xia2022less}
{\sc Xia, C.~S., and Zhang, L.}
\newblock Less training, more repairing please: revisiting automated program repair via zero-shot learning.
\newblock In {\em Proceedings of the 30th ACM Joint European Software Engineering Conference and Symposium on the Foundations of Software Engineering\/} (2022), pp.~959--971.

\bibitem{ye2021comfort}
{\sc Ye, G., Tang, Z., Tan, S.~H., Huang, S., Fang, D., Sun, X., Bian, L., Wang, H., and Wang, Z.}
\newblock Automated conformance testing for javascript engines via deep compiler fuzzing.
\newblock In {\em Proceedings of the 42nd ACM SIGPLAN International Conference on Programming Language Design and Implementation\/} (2021), ACM, pp.~435--450.

\bibitem{ziegler2022productivity}
{\sc Ziegler, A., Kalliamvakou, E., Li, X.~A., Rice, A., Rifkin, D., Simister, S., Sittampalam, G., and Aftandilian, E.}
\newblock Productivity assessment of neural code completion.
\newblock In {\em Proceedings of the 6th ACM SIGPLAN International Symposium on Machine Programming\/} (2022), pp.~21--29.

\end{thebibliography}

\end{document}